# Localized orbitals for optimal decomposition of molecular properties

T.Yu.Nikolaienko[*] and L.A.Bulavin

*Faculty of Physics of Taras Shevchenko National University of Kyiv;*
*64/13, Volodymyrska Street, 01601 Kyiv, Ukraine.*

**Abstract**
Localized orbitals are important for modeling and interpreting complicated electronic structures of atoms and molecules in a chemically meaningful way. Here, we present the parameter-free procedure for transforming delocalized molecular orbitals (either canonical self-consistent field orbitals, or Lowdin natural orbitals obtained from a general wavefunction) into the localized property-optimized orbitals (LPOs), which can be used for building the most accurate (in the Frobenius norm sense) approximation to the first-order reduced density matrix in form of the sum of localized mono- and diatomic terms. In this way any, any one-electron molecular property can be decomposed into contributions associated with individual atoms and the pairs of atoms, with the upper bound for the decomposition acucracy known in advance due to Cauchy–Bunyakovsky–Schwarz inequality. In addition, an algorithm is proposed for obtaining 'the Chemist's LPOs' (CLPOs) set of localized orbitals containing a single orbital per a pair of electrons and forming an idealized Lewis structure with the one-electron properties which are closest to the properties obtained from the original many-electron wavefunction. The computational algorithms for constructing LPOs and CLPOs as well as their underlying atomic hybrid orbitals (AHOs and LHOs respectively) from the results of quantum-chemical calculations are presented and their implementation within the open-source freeware program JANPA (http://janpa.sourceforge.net/ ) is discussed. The performance of the proposed orbital localization procedures is assessed using the test set of density matrices of 33432 small molecules obtained at Hartree-Fock and 2-nd order Moller-Plesset theory levels.

## 1. Introduction

Localized molecular orbitals resulting from a unitary transformation of occupied canonical molecular orbitals[1] (MOs) play essential role in physical chemistry as the 'building blocks' or 'descriptors' through which the complicated electronic structure of atoms and molecules can be modeled or interpreted in a comprehensible way. In addition, the localized orbitals concentrated in a limited spatial region of a molecule proved useful in making the high-level correlated quantum-chemical methods more computationally tractable[2–8]. Although the concept of an orbital itself has been much methodologically debatable[9–33] and is too 'fuzzy'[34] to be defined more precisely than just as the function of coordinates of a single electron *somehow* related to the system's wavefunction or electron density, this concept still remains virtually the best one proposed so far for moving the ideas of valence electrons and electron pairs (including bonding and antibonding orbitals, lone pairs etc.), which are the key elements of the chemist's Lewis-structure picture[35], from qualititative to a quantum-mechanical ground. From this perspective, the localized orbitals are used to decompose (typically in an approximate manner) true many-electron wavefunction of the molecule into the components allowing a chemically meaningful interpretation.

In order to transform delocalized orbitals (either the canonical MOs obtained as a solutions of self-consistent field Hartree-Fock or Kohn-Sham equations[36], or the Lowding natural orbitals[37] obtained by diagonalization of the first-order density matrix corresponding to a correlated wavefunction) into the localized ones, a number of procedures have been developed[4,38–55]. However, most of them date back to the times preceding the 'mini revolution' of 1970's[56], when personal computers became widely spread among chemists, and therefore these procedures were implemented as a handy computer codes, with probably the only exception of the Natural Bond Orbital (NBO) method[38–44] implemented in NBO program[44]. Moreover, most of the localization procedures developed so far focus on optimizing the localized orbitals for representing only one quantity, e.g., electron charge[41,42], local orbital populations[49], bond order[50,51], Coulomb repulsion[52], correlation energy contributions[4,53,54] etc.

---

[*] Author to whom correspondence should be addressed: tim_mail [that_typical_symbol_in_addresses] ukr.net

In this paper we propose a new orbital localization procedure, along with its implementation in the open-source program JANPA[57,58], for obtaining the localized orbitals (LPOs) designed to be equally suitable for optimal decomposition of any of one-electron properties into mono- and diatomic contributions. This optimality is ensured by defining the LPOs $\varphi^{loc}_i(\vec{r})$ as a set of functions, with each function concentrated at a single atom or a pair of atoms, which provides the most accurate (in a Frobenius norm sense) localized approximation

$$\gamma^{loc}(\vec{r},\vec{r}') = \sum_i n^{loc}_i \cdot \varphi^{loc}_i(\vec{r})\varphi^{loc}_i(\vec{r}') \tag{1}$$

for spin independent (spin traced) first-order reduced density matrix (1-RDM)[59–61]

$$\gamma(\vec{r},\vec{r}') = N_e \cdot \sum_{\sigma_1...\sigma_N} \int \Psi^{*\sigma_1...\sigma_N}(\vec{r},\vec{r}_2,...,\vec{r}_N) \cdot \Psi^{\sigma_1...\sigma_N}(\vec{r}',\vec{r}_2,...,\vec{r}_N) \cdot d\vec{r}_2...d\vec{r}_N, \tag{2}$$

where $\Psi$ is the molecular wavefunction, $N_e$ is the total number of electrons and $\sigma_i$ stands for electron spin indices. It is essential that 1-RDM, as defined by Eq. (2), contains all the necessary information about the electronic structure of the system, which is rquired to compute any of the system (spin-independent) one-electron properties[62] defined by the quantum-mechanical operator

$$\hat{F} = \sum_{i=1}^{N_e} \hat{f}(\vec{r}_i)$$

as the expectation value

$$\langle \hat{F} \rangle = \int \left( \hat{f}(\vec{r}) \gamma(\vec{r},\vec{r}') \right)_{\vec{r}'=\vec{r}} d\vec{r} = \mathrm{tr}\left( \hat{f}\hat{\gamma} \right), \tag{3}$$

where $\hat{\gamma}$ is the first-order reduced density operator[35,62] (1-RDO) defined as the operator posessing 1-RDM as the kernel. In addition, many of the chemically meaningful electronic structure descriptors, such as effective atomic charges[42,57], bond orders and valencies[62–67], can be obtained from the information entirely contained in 1-RDM. Many more characteristics of the system can also be obtained from 1-RDM through more complicated non-linear and/or implicit relationships taking into account that 1-RDM 'diagonal part' $\gamma(\vec{r},\vec{r})$ equals the system's electron density $\rho(\vec{r})$ which determines virtually all the physical properties of a many-electron system in its ground state through the density functional theory (DFT) framework[36].

Now, if the true 1-RDO $\hat{\gamma}$ in (3) is substituted by its localized approximation $\hat{\gamma}^{loc}$, defined by (1) as the kernel, we arrive at the corresponding localized decomposition

$$\langle \hat{F} \rangle^{loc} = \mathrm{tr}\left( \hat{f}\hat{\gamma}^{loc} \right) = \sum_i n_i^{loc} \langle \varphi_i^{loc} | \hat{f} | \varphi_i^{loc} \rangle = \sum_i n_i^{loc} F_i^{loc}$$

of molecular property $\hat{F}$ into localized contributions $F_i^{loc} = \langle \varphi_i^{loc} | \hat{f} | \varphi_i^{loc} \rangle$, each being related to a single localized orbital $\varphi_i^{loc}$, which in turn, can be attributed to the contribution from a certain atom or atomic pair. Although $\hat{\gamma}^{loc}$ is only an approximation to the true $\hat{\gamma}$, and hence $\langle \hat{F} \rangle^{loc}$ equals the true expectation value $\langle \hat{F} \rangle$ only approximately, the discrepancy between the last two quantities is minimized as soon as difference $\|\gamma - \gamma^{loc}\|^2$ is minimized, as can be justified by the trace version of Cauchy–Bunyakovsky–Schwarz inequality[68] implying

$$\left| \langle \hat{F} \rangle^{loc} - \langle \hat{F} \rangle \right| = \left| \mathrm{tr}\left( \hat{f}\left( \hat{\gamma}^{loc} - \hat{\gamma} \right) \right) \right| \leq \sqrt{\|\hat{\gamma}^{loc} - \hat{\gamma}\|^2} \cdot \sqrt{\|\hat{f}\|^2}.$$

From this inequality, $\|\hat{\gamma}^{loc} - \hat{\gamma}\|^2$ becomes a natural choice for the criteria to be applied in the search for the localized orbitals optimally suitable for decomposition of any of the one-electron properties into localized contributions. Due to this fact, we shall hereinafter refer to the localized orbitals $\varphi^{loc}_i(\vec{r})$ which minimize $\left\|\hat{\gamma}^{loc} - \hat{\gamma}\right\|^2$ as the localized property-optimal orbitals (LPOs). In the following sections we develop the procedure for obtaining LPOs from 1-RDM expressed in terms of atomic basis functions and analyze the procedure performance on the large test set of 33432 small molecules containing from 2 to 12 atoms. For simplicity only a closed-shell case is considered throughout the paper.

## 2. Theory

*2.1. Localization criterion and constraints on the localized orbital composition*

We now turn to the development of the method for building such LPOs $\varphi^{loc}_i(\vec{r})$ and their corresponding coefficients $n^{loc}_i$ (see (1)) which minimize

$$\left\|\hat{\gamma}^{loc} - \hat{\gamma}\right\|^2 = \int \left(\sum_i n^{loc}_i \cdot \varphi^{loc}_i(\vec{r})\varphi^{loc}_i(\vec{r}') - \gamma(\vec{r},\vec{r}')\right)^2 d\vec{r}d\vec{r}' \to \min_{n^{loc}_i, \varphi^{loc}_i(\vec{r})}. \qquad (4)$$

It is convenient first to find the optimal values of $n^{loc}_i$ and then proceed with searching for the localized orbitals $\varphi^{loc}_i(\vec{r})$. We will assume the latter to be real-valued functions, normalized to unity and orthogonal, i.e.,

$$\left(\varphi^{loc}_i, \varphi^{loc}_j\right) = \int \varphi^{loc}_i(\vec{r}) \cdot \varphi^{loc}_j(\vec{r}) = \delta_{ij}$$

where $\delta_{ij}$ is the Kronecker delta. Given this property, the target criteria (4) can be re-expressed as

$$\left\|\hat{\gamma}^{loc} - \hat{\gamma}\right\|^2 = \sum_i \left(n^{loc}_i\right)^2 - 2\sum_i n^{loc}_i \cdot \int \varphi^{loc}_i(\vec{r})\gamma(\vec{r},\vec{r}')\varphi^{loc}_i(\vec{r}')d\vec{r}d\vec{r}' + \left\|\hat{\gamma}\right\|^2, \qquad (5)$$

where $\left\|\hat{\gamma}\right\|^2$ is independent of both $\varphi^{loc}_i(\vec{r})$ and $n^{loc}_i$. The minimization of the obtained expression as the function of $n^{loc}_i$ gives

$$n^{loc}_i = \int \varphi^{loc}_i(\vec{r})\gamma(\vec{r},\vec{r}')\varphi^{loc}_i(\vec{r}')d\vec{r}d\vec{r}' = \left(\varphi^{loc}_i, \hat{\gamma}\varphi^{loc}_i\right), \qquad (6)$$

so that the coefficients $n^{loc}_i$ appear as the LPO occupancies.

It should be noted that if (5) is minimized with respect to $\varphi^{loc}_i(\vec{r})$ without any further constraints, this results in a well-known Löwdin natural orbitals[37] $\varphi^{nat.}_i(\vec{r})$, which allow for exact ($\left\|\hat{\gamma}^{nat.} - \hat{\gamma}\right\|^2 = 0$) expansion of the true 1-RDM as

$$\gamma^{nat.}(\vec{r},\vec{r}') = \sum_i n^{nat.}_i \cdot \varphi^{nat.}_i(\vec{r})\varphi^{nat.}_i(\vec{r}'). \qquad (7)$$

Clearly, the equality $\left\|\hat{\gamma}^{nat.} - \hat{\gamma}\right\|^2 = 0$ holds strictly only if the infinite number of terms is included into sum (7). However, since in practical quantum-chemical calculations the wavefunction and/or its components are always presented in the form of expansion over a finite number of basis functions, it can be also taken that $\left\|\hat{\gamma}^{nat.} - \hat{\gamma}\right\|^2 = 0$ remains true for 1-RDM obtained from such approximate calculations provided that the number of natural orbitals equals the number of original (linearly independent) basis functions used in the calculations. Still, the expansion (7) requires the use of orbitals which are typically delocalized over the entire molecule, just like the canonical MOs. Therefore, in order to obtain localized orbitalsm certain constraints must inevitably be imposed on $\varphi^{loc}_i$.

In the proposed method such constraints can be best expressed in terms of coefficients of expansions

$$\varphi^{loc}_{i}(\vec{r}) = \sum_{\mu} c^{loc}_{i\mu} \chi_{\mu}(\vec{r}) \tag{8}$$

of localized $\varphi^{loc}_{i}$ and

$$\varphi^{nat.}_{i}(\vec{r}) = \sum_{\mu} c^{nat.}_{i\mu} \chi_{\mu}(\vec{r})$$

of natural (or canonical) orbitals $\varphi^{nat.}_{i}(\vec{r})$ over the set of atom-centered basis functions which are assumed to be real-valued, normalized to unity and orthogonal (in both intra- and interatomic sense), i.e., $(\chi_{\mu}, \chi_{\nu}) = \delta_{\mu\nu}$. For the sake of brevity, we shall further refer to these functions as AOs. Upon introducing AOs, each of which can be attributed to a particular atom in the system, the localization of the orbital is understood as the property of expansion (8) to contain only the AOs centered at a single atom or at any of the two atoms, hereinafter referred to as an atomic pair. Applying this restriction, Eq. (8) can be rewritten as

$$\varphi^{loc}_{i}(\vec{r}) = \sum_{\mu \in X} c^{loc}_{i\mu} \chi_{\mu}(\vec{r}),$$

where $X$ denotes a limited set of values the summation index can take.

We further introduce a specific additional constraint[38] on the structure of desired localized orbitals. There should exist such orthonormal linear combinations, further on referred to as atomic hybrid orbitals (AHOs) or simply 'hybrids',

$$h_{i}^{(A)}(\vec{r}) = \sum_{\mu \in A} \Theta_{\mu i} \cdot \chi_{\mu}(\vec{r}) \tag{9}$$

of the AOs $\chi_{\mu}$ on each atom that any two-center orbital localized at the atomic pair A–B (typically representing a covalent bond between the atoms) are expressible as

$$\varphi^{loc}_{i}(\vec{r}) = \sum_{\mu \in A \cup B} c^{loc}_{i\mu} \chi_{\mu}(\vec{r}) = v_{i,A} h_{i}^{(A)}(\vec{r}) + v_{i,B} h_{i}^{(B)}(\vec{r}), \tag{10}$$

where $v_{i,A}^{2} + v_{i,B}^{2} = 1$ and neither of these coefficients is close to zero while any orbital localized at a single atom $M$ (typically representing its vacant orbital or an unshared electron pair) are expressible as

$$\varphi^{loc}_{j}(\vec{r}) = \sum_{\mu \in M} c^{loc}_{j\mu} \chi_{\mu}(\vec{r}) = h_{j}^{(M)}(\vec{r}). \tag{11}$$

We shall refer to these two types of localized orbitals as 2c-LOs and 1c-LOs, respectively, for the sake of brevity. The number of orthonormal AHOs on each particular atom is mainained equal to the number of original AOs for the same atom, and both sets of functions span the same Hilbert space. We also require that the total number of all LOs (2c-LOs and 1c-LOs) shoule be the same as the number of AHOs (and hence, of AOs) and that they too should span the same Hilbert space. These requirements guarantee the existence of orthogonal transformations which relate the sets of AOs, AHOs and LOs to one another.

*2.2. Criterion for optimal atomic hybrid orbitals*

The above constraint on the structure of LOs has important implications. Together with the requirement of orthonormality of the LOs it implies that if the AHO enters a certain 1c-LO, such AHO must be orthogonal to all the other LOs (either 1c-LOs or 2c-LOs) and hence can not appear in the (10)- or (11)-type expansion of any of them. Likewise, if the AHO enters 2c-LO, such AHO can not appear in the (11)-type expansion of any 1c-LO (since otherwise some LOs won't be orthogonal), but can additionally appear in the (10)-type expansion of only a single additional 2c-LO at most. What is more, since the total number of LOs equals the total number of AHOs, we conclude that if the AHO enters a certain 2c-LO, it must also appear in exactly a single additional 2c-LO. Consequently, under appropriate orbital numbering scheme the AHOs appearing in 2c-LOs (and not appearing in any of 1c-LOs) can be grouped into pairs in such a way that there would be a one-to-one correspondence between the pair of AHOs $\{h_i(\vec{r}), h_{p(i)}(\vec{r})\}$ and the pair of 2c-LOs $\{\varphi^{loc}_i(\vec{r}), \varphi^{loc}_{p(i)}(\vec{r})\}$. Here we have introduced a discrete 'pairing function' $p(i)$, which is involutive ($p(p(i)) = i$) and equals to the index of the 'partner' AHO belonging to the same pair as the $i$-th AHO. Then linear relations (10) between the members of the pairs can be expressed in the matrix form

$$\boldsymbol{\varphi}^{loc}_i(\vec{r}) = \mathbf{v}_i \mathbf{h}_i(\vec{r}), \qquad (12)$$

in which we have introduced column-vectors

$$\boldsymbol{\varphi}^{loc}_i(\vec{r}) = \begin{pmatrix} \varphi^{loc}_i(\vec{r}) \\ \varphi^{loc}_{p(i)}(\vec{r}) \end{pmatrix}$$

and

$$\mathbf{h}_i(\vec{r}) = \begin{pmatrix} h_i(\vec{r}) \\ h_{p(i)}(\vec{r}) \end{pmatrix},$$

containing 2c-LOs and AHOs of the corresponding pairs, as well as a 2x2 matrix $\mathbf{v}_k$ containing the AHO 'mixing coefficients'. As can be concluded from the orthonormality conditions introduced above for both AHOs and 2c-LOs, the matrix $\mathbf{v}_k$ must be orthogonal, i.e., $\mathbf{v}_k^T \mathbf{v}_k = \mathbf{v}_k \mathbf{v}_k^T = \mathbf{1}$.

To summarize, the constraints (10) and (11) imposed on the structure of LOs result in splitting the entire set of AHOs into two subsets:

a) the '1C subset' which consists of AHOs identical to 1c-LOs in accordance with (11) and contains as many functions as there are 1c-LOs,

b) the '2C subset' which consists of the pairs of AHOs $\mathbf{h}_k$, each pair being in one-to-one correspondence with a pair of 2c-LOs $\boldsymbol{\varphi}^{loc}_k$ via orthogonal transformation (12). Below we shall use notation $i \in 2C$ to indicate that there exists a pair of AHOs in 2C subset to which the $i$-th AHO belongs.

The deduced restriction on the number of LOs into which each of the AHOs can enter leads readily to the criterion for finding AHOs. Indeed, since AHOs span the same Hilbert space as the original AOs, the AHOs can be used as the basis for expanding the true 1-RDM over it. It is thus possible to introduce the density matrix $\mathbf{D}$ by defining its elements as the expansion coefficients

$$\gamma(\vec{r},\vec{r}') = \sum_{\mu,\nu} (\mathbf{D})_{\mu\nu} \cdot h_\mu(\vec{r}) h_\nu(\vec{r}'), \qquad (13)$$

where we have dropped for simplicity the superscripts at AHOs. Due to orthonormality of AHOs we also conclude that the elements of $\mathbf{D}$ can be obtained as

$$(\mathbf{D})_{\mu\nu} = (h_\mu, \hat{\gamma} h_\nu). \qquad (14)$$

Similarly to (13), the localized approximate 1-DRM $\gamma^{loc}$ can be expanded over the same set of basis functions as

$$\gamma^{loc}(\vec{r},\vec{r}') = \sum_{\mu,\nu} (\mathbf{D}^{loc})_{\mu\nu} \cdot h_\mu(\vec{r}) h_\nu(\vec{r}'), \qquad (15)$$

thereby defining the elements $(\mathbf{D}^{loc})_{\mu\nu}$ of another matrix $\mathbf{D}^{loc}$. Similarly to (14), we find that $(\mathbf{D}^{loc})_{\mu\nu} = (h_\mu, \hat{\gamma}^{loc} h_\nu)$, and employing (1) we also find that

$$(\mathbf{D}^{loc})_{\mu\nu} = \sum_i n^{loc}_i \cdot (\varphi^{loc}_i, h_\mu)(\varphi^{loc}_i, h_\nu).$$

It follows from this expression that, given the properties of LOs $\varphi^{loc}_i$ summarized by (10) and (11), only the following elements of $\mathbf{D}^{loc}$ can be non-zero: all of its diagonal elements and those (and only those) off-diagonal elements which correspond to the AHOs belonging to the same pair in the '2C subset'. What is more, since

$$\left\|\hat{\gamma}^{loc} - \hat{\gamma}\right\|^2 = \left\|\mathbf{D}^{loc} - \mathbf{D}\right\|^2 = \sum_{\mu,\nu}\left(\left(\mathbf{D}^{loc}\right)_{\mu\nu} - \left(\mathbf{D}\right)_{\mu\nu}\right)^2, \quad (16)$$

all of the non-zero elements of $\mathbf{D}^{loc}$ must be set equal to the corresponding elements of $\mathbf{D}$ in order to minimize $\left\|\hat{\gamma}^{loc} - \hat{\gamma}\right\|$, thus making $\gamma^{loc}$ as close to $\gamma$ as possible, which leads to

$$\left(\mathbf{D}^{loc}\right)_{\mu\nu} = \begin{cases} \left(\mathbf{D}\right)_{\mu\mu}, & \text{if } \mu = \nu \\ \left(\mathbf{D}\right)_{\mu\nu}, & \text{if } \mu \in 2C \text{ and } \nu = p(\mu), \\ 0, & \text{otherwise} \end{cases} \quad (17)$$

where the condition $\nu = p(\mu)$ indicates that the $\mu$-th and $\nu$-th AHOs both belong to the same pair in the 2C subset. Given (17), we proceed with the approximation error (16) as

$$\left\|\mathbf{D}^{loc} - \mathbf{D}\right\|^2 = \left\|\mathbf{D}\right\|^2 - \sum_{\mu}\left(\left(\mathbf{D}\right)_{\mu\mu}\right)^2 - \sum_{\mu \in 2C}\left(\left(\mathbf{D}\right)_{\mu,p(\mu)}\right)^2.$$

It should be noted that each term involving the elements of symmetrical matrix $\mathbf{D}$ appears twice in the summation over $\mu \in 2C$: first in the form of $\left(\left(\mathbf{D}\right)_{\mu,p(\mu)}\right)^2$ and then as $\left(\left(\mathbf{D}\right)_{p(\nu),\nu}\right)^2$ when the index $\mu$ coincides with $p(\nu)$ for a certain value of $\nu \in 2C$.

Finally, since AHOs are merely the result of orthogonal transfromation (9) of original AOs and the sum of squared elements of the matrix ($\left\|\mathbf{D}\right\|^2 = \sum_{\mu,\nu}\left(\left(\mathbf{D}\right)_{\mu\nu}\right)^2 = inv.$) is invariant under such transformations, it can be concluded that $\left\|\mathbf{D}^{loc} - \mathbf{D}\right\|^2$ is minimized as soon as the sum of squared non-zero elements (17) of $\mathbf{D}^{loc}$ is maximized, i.e.,

$$\Phi_{AHO} = \sum_{\mu}\left(\left(\mathbf{D}\right)_{\mu\mu}\right)^2 + \sum_{\mu \in 2C}\left(\left(\mathbf{D}\right)_{\mu,p(\mu)}\right)^2 = \sum_{\mu}\left(h_\mu,\hat{\gamma}h_\mu\right)^2 + \sum_{\mu \in 2C}\left(h_\mu,\hat{\gamma}h_{p(\mu)}\right)^2 \to \max. \quad (18)$$

This condition comprises the guiding principle in the search of AHOs.

*2.3. The AHO optimization algorithm*

Condition (18) suggests that the procedure for finding AHOs should be performed by ensuring that:
    i) the unitary transformation (9) is built so as to make the elements of $\mathbf{D}$ appearing in (18) as large by their absolute values as possible,
    ii) the AHOs are grouped into pairs, thus forming the 2C subsets, in such a way that the second sum in Eq. (18) is as large as possible ('an optimal pairing requirement').

These two requirements are clearly interdependent and in general can be best satisfied in an iterative manner with each iteration involving two sub-steps. The first sub-step consists in building the AO-to-AHO transformation (9) and is executed differently at the very first iteration and at all subsequent iterations.

At the very first iteration of our implementation the requirement i) is fulfilled approximately by simultaneous diagonalization algorithm (SDA)[69,70] operating on the set matrices

$$G_A = \left\{\tilde{\mathbf{D}}_{AA}^T \tilde{\mathbf{D}}_{AA}\right\} \cup \left\{\tilde{\mathbf{D}}_{AB}\tilde{\mathbf{D}}_{BA}, B \neq A\right\},$$

defined individually for each atom A, where $\tilde{\mathbf{D}}_{AB} = \tilde{\mathbf{D}}_{BA}^T$ denotes the sub-matrix of $\mathbf{D}$ composed of elements $\left(\mathbf{D}\right)_{\mu\nu}$ with indices $\mu \in A, \nu \in B$. The SDA finds the unitary transformations $\Theta_A$ in the space of AOs of each atom, which minimizes the sum of squared off-diagonal components

$$\mathrm{off}(A) = \sum_{\nu} \sum_{\mu \neq \nu} \left( \left( \tilde{\mathbf{D}}_{AA}^T \tilde{\mathbf{D}}_{AA} \right)_{\mu\nu} \right)^2 + \sum_{B \neq A} \sum_{\nu} \sum_{\mu \neq \nu} \left( \left( \tilde{\mathbf{D}}_{AB} \tilde{\mathbf{D}}_{BA} \right)_{\mu\nu} \right)^2 ,$$

or equivalently, maximizes the sum of squared diagonal components

$$\mathrm{dg}(A) = \sum_{\nu} \left( \left( \tilde{\mathbf{D}}_{AA}^T \tilde{\mathbf{D}}_{AA} \right)_{\nu\nu} \right)^2 + \sum_{B \neq A} \sum_{\nu} \left( \left( \tilde{\mathbf{D}}_{AB} \tilde{\mathbf{D}}_{BA} \right)_{\nu\nu} \right)^2 . \qquad (19)$$

Note that only two approximations are involved at this initial iteration: the SDA target function (19) contains a greater number of the matrix **D** elements than the ultimate AHO target function (18), and the matrix **D** elements to the 4th rather than to the 2nd power are present in (19).

The set of $\boldsymbol{\Theta}_A$ matrices obtained by SDA provides the initial approximation for AO-to-AHO transformations, which make most of the elements of **D** in AHO basis as small as possible and at the same time maximize the few remaining elements of **D** on the diagonals of $\tilde{\mathbf{D}}_{AA}$ and $\tilde{\mathbf{D}}_{AB}$ sub-matrices. After the transformation of this kind the elements of **D** become suitable for combining them into pairs to satisfy the optimal pairing requirement ii).

This requirement is satisfied at the second sub-step of each iteration. During this sub-step, which is executed in the same way at all iterations, the AO-to-AHO transformation (9) is assumed fixed, so that the first term in (18) is constant and the optimal pairing requirement becomes essentially identical to the maximum bond order principle by Jug[50,51]. Note, however, that in our approach this principle is a consequence of the more general optimality condition (4) subjected to restrictions (10)–(11) on the structure of localized orbitals, and therefore it should not be considered as an independent postulate.

The optimal pairing requirement is fulfilled in our implementation by employing the maximum-weight matching 'blossom algorithm' by J. Edmonds[71,72]. This algorithm finds a set of edges connecting each of the given nodes numbered from 1 to $n$ to at most another single node in such way that the created matching would maximize the sum of weights corresponding to the edges. More formally, if a weight $w_{ij}$, satisfying $w_{ij} = w_{ji}$ and $w_{ii} = 0$, is associated with the edge connecting the $i$-th and $j$-th nodes, and symmetrical $n \times n$ adjacency matrix **P** is introduced so as $P_{ij} = 1$ if and only if there is an edge connecting the $i$-th and $j$-th nodes (in particlar, $P_{ii} = 1$ if the $i$-th node is *not* connected to any of other node) and $P_{ij} = 0$ otherwise, the blossom algorithm finds the adjacency matrix which maximizes the cost function $\sum_{i,j} w_{ij} P_{ij}$ under constraints $\sum_{j} P_{ij} \in \{0,1\}$ for all $i$ from 1 to $n$ (which indicates that each node can have no more than a single 'partner'). The algorithm has polynomial time complexity (namely $O(n^3)$ in our code, with $n$ being the total number of AHOs), and is executed with the weights $w_{\mu\nu}$ set to $\left( (\mathbf{D})_{\mu\nu} \right)^2$, where the elements of matrix **D** are computed by (14) using the AO-to-AHO transformation matrices available at the current iteration. Note that an optimal distribution of AHOs into 1C subset and pairs of 2C subset is found from the obtained adjacency matrix **P** by a simple rule: the $i$-th AHO is placed into 1C subset if $P_{ii} = 1$ and is placed into 2C subset paired with the $j$-th AHO if $P_{ij} = 1$. The elements of the adjacency matrix are related to the above pairing function $p(i)$, defined on elements of 2C subset, as follows: $p(i) = j$ where $i \neq j$ if and only if $P_{ij} = 1$.

After the optimal pairing has been found, the next iteration begins. At this, as well as at all further iterations, the current approximations to both the AO-to-AHO transformation and the optimal pairing function are available. Now the sub-step i) is executed differently and is aimed at improving the unitary AO-to-AHO transformation for maximizing $\Phi_{AHO}$ defined in (18) under assumption that the AHO pairing is fixed. In the current implementation this optimization is performed by a simple iterative algorithm, in a certain way similar to the steepest descent algorithm, with orthogonality constraint (cf. ref. 73 and references therein). The algorithm is described in details in Appendix B and is based on the following reasoning.

Consider the optimization problem of finding an orthogonal matrix $\mathbf{U}$ ($\mathbf{U}^T\mathbf{U} = \mathbf{1}$) which maximizes some differentiable target function $f(\mathbf{U})$. If this function is linear, i.e.,

$$f(\mathbf{U}) = \text{tr}(\mathbf{G}^T\mathbf{U}) + f_0, \tag{20}$$

where $f_0$ is constant, the unitary matrix $\mathbf{U}$, for which $f$ achieves its maximum value among all orthogonal matrices of a given dimension, can be found as

$$\mathbf{U} = \mathbf{G}(\mathbf{G}^T\mathbf{G})^{-1/2}. \tag{21}$$

This result is a direct consequence of the fact that such $\mathbf{U}$ provides a unitary matrix closest (in the Frobenius norm sense) to the given (non-unitary) matrix $\mathbf{G}$, i.e., $\mathbf{U}$ minimizes $\|\mathbf{G} - \mathbf{U}\|^2 = \mathbf{G}^T\mathbf{G} - 2\,\text{tr}(\mathbf{G}^T\mathbf{U}) + 1$.[74–76] Note that the same matrix $\mathbf{U}$ can be alternatively computed from $\mathbf{G}$ by finding its singular value decomposition (SVD)

$$\mathbf{G} = \mathbf{V}\boldsymbol{\Lambda}\mathbf{Q}, \tag{22}$$

where $\mathbf{V}$ and $\mathbf{Q}$ are unitary matrices and $\boldsymbol{\Lambda}$ is a diagonal one. Indeed, the symmetrical square root of $\mathbf{G}^T\mathbf{G} = \mathbf{Q}^T\boldsymbol{\Lambda}^2\mathbf{Q}$ can be found as[77] $(\mathbf{G}^T\mathbf{G})^{1/2} = \mathbf{Q}^T\boldsymbol{\Lambda}\mathbf{Q}$ and its inverse as $(\mathbf{G}^T\mathbf{G})^{-1/2} = \mathbf{Q}^T\boldsymbol{\Lambda}^{-1}\mathbf{Q}$. Finally, the substitution of the latter result together with (22) into (21) yields[78]

$$\mathbf{U} = \mathbf{V}\boldsymbol{\Lambda}\mathbf{Q}\mathbf{Q}^T\boldsymbol{\Lambda}^{-1}\mathbf{Q} = \mathbf{V}\mathbf{Q}. \tag{23}$$

This reformulation allows avoiding the possible numerical instabilities in computing the inverse square root $(\mathbf{G}^T\mathbf{G})^{-1/2}$ in case if $\mathbf{G}^T\mathbf{G}$ has small (or nearly zero) eigenvalues.

Now assume that the target function $f$ depends on the components of matrix $\mathbf{U}$ in non-linear manner, which is the case in (18) for the dependence of the target function $\Phi_{AHO}$ on the coefficients of AO-to-AHO transformation. In this case matrix $\mathbf{U}$ can be found as follows.

Given that the AO-to-AHO transformation optimization algorithm is initialized with the result of SDA, which maximizes a similar target function (19), it can be anticipated that the ultimate transformation will appear rather 'close' to its initial guess. It is therefore reasonable to assume that due to the initialization used in our implementation, the target function can be approximated by the linear function quite accurately in the vicinity of an initial guess $\mathbf{U}_0$. We thus consider the first-order expansion of the target function $f$ into the Taylor series in the neighborhood of $\mathbf{U}_0$:

$$f(\mathbf{U}) \approx f(\mathbf{U}_0) + \sum_{\mu,\nu} \left.\frac{\partial f}{\partial(\mathbf{U})_{\mu\nu}}\right|_{\mathbf{U}=\mathbf{U}_0} (\mathbf{U}-\mathbf{U}_0)_{\mu\nu} = f(\mathbf{U}_0) + \text{tr}(\mathbf{G}^T(\mathbf{U}-\mathbf{U}_0)) = F(\mathbf{U}),$$

where $\mathbf{G}$ is the 'gradient matrix' with the elements

$$(\mathbf{G})_{\mu\nu} = \left.\frac{\partial f}{\partial(\mathbf{U})_{\mu\nu}}\right|_{\mathbf{U}=\mathbf{U}_0}. \tag{24}$$

Since $\mathbf{U} = \mathbf{G}(\mathbf{G}^T\mathbf{G})^{-1/2}$ rather than $\mathbf{U} = \mathbf{U}_0$ maximizes $\text{tr}(\mathbf{G}^T\mathbf{U})$, we conclude that $\text{tr}(\mathbf{G}^T\mathbf{U}) \geq \text{tr}(\mathbf{G}^T\mathbf{U}_0)$, so that $\text{tr}(\mathbf{G}^T(\mathbf{U}-\mathbf{U}_0)) \geq 0$ and the hence $F(\mathbf{U}) > F(\mathbf{U}_0)$.

In case if the linear approximation $f(\mathbf{U}) \approx F(\mathbf{U})$ were strictly guaranteed to be accurate, implying that $f(\mathbf{U}) \geq f(\mathbf{U}_0)$, it would be sufficient to repeatedly update the gradient matrix according to (24) and then re-compute the new matrix $\mathbf{U}$ according to (21) or (23) thus establishing an iterative procedure for finding $\mathbf{U}$ which maximizes $f$ (cf. ref. 79). In this case it would be sufficient to require that the target function $f$ is bounded from above in order to guarantee that this iterative procedure will converge. Given that the target function (18) of the AHO optimization problem is bounded from above at least as $\Phi_{AHO} \leq \|\mathbf{D}\|^2$, the above argumentation for the convergence criteria of the iterative procedure would still hold provided that inequality
$$f(\mathbf{U}) \geq f(\mathbf{U}_0) \qquad (25)$$
holds at each iteration. Therefore, this inequality must be checked numerically at each iteration.

If inequality (25) is violated at some step, i.e., $f(\mathbf{U}) < f(\mathbf{U}_0)$ for the new $\mathbf{U}$ obtained from (21) or (23), this indicates that the linear approximation $f(\mathbf{U}) \approx F(\mathbf{U})$ has failed because of too large change of the argument $\mathbf{U}$ relative to its preceeding value $\mathbf{U}_0$. In this case it is possible to adjust the 'stepsize' by modifying Eq. (21) in such a way that inequality $f(\mathbf{U}) \geq f(\mathbf{U}_0)$ becomes valid again and the convergence argumentation outlined above is restored. As shown in Appendix C, such adjustment of the step size can be achieved by adding a penalty term to the target function, ultimately leading to the replacement of the gradient matrix defined in (24) by the 'damped gradient'
$$\tilde{\mathbf{G}} = \lambda \cdot \mathbf{G} + \mathbf{U}_0$$
and then using it in Eq. (21) or (23) instead of the true gradient matrix $\mathbf{G}$ to compute a new $\mathbf{U}$. The proposed procedure for choosing the value of the 'damping parameter' $\lambda$ is described in Appendix C.

Generalization onto the case of a target function $f(\mathbf{U}_1,...,\mathbf{U}_{N_a})$ depending on $N_a$ matrix arguments is straightforward: the first-order expansion is represented as
$$f(\mathbf{U}_1,...,\mathbf{U}_{N_a}) \approx f(\mathbf{U}_1^{(0)},...,\mathbf{U}_a^{(0)}) + \sum_k \mathrm{tr}\left(\mathbf{G}_k^T (\mathbf{U}_k - \mathbf{U}_k^{(0)})\right),$$
where the partial gradient matrices $\mathbf{G}_k$ ($k = 1,2,..,N_a$) have the elements
$$(\mathbf{G}_k)_{\mu\nu} = \left.\frac{\partial f}{\partial (\mathbf{U}_k)_{\mu\nu}}\right|_{\mathbf{U}_l = \mathbf{U}_l^{(0)}}.$$
In order to treat all arguments of $f$ on equal footing, the new values of all arguments must be computed at once, after the corresponding matrices $\mathbf{G}_k$ have been obtained using the 'old' argument values $\mathbf{U}_1^{(0)},...,\mathbf{U}_a^{(0)}$. Inequality (25) must be then checked and the usual stepsize adjustment must be executed in case if (25) is violated.

After these preparations, it is now possible to formulate the complete AHO construction algorithm:

*Step 1.* Initialize AO-to-AHO transformation by simultaneous diagonalization algorithm and compute $\mathbf{D}$.

*Step 2.* Determine an optimal hybrid pairing and create the 2C subset.

*Step 3.* Optimize AO-to-AHO transformation by the algorithm from Appendix B with convergence threshold $\theta_{occh}$ and compute the target function using the currently available 2C subset.

*Step 4.* Compute $\mathbf{D}$ using optimized AO-to-AHO transformation, then determine the optimal hybrid pairing and calculate the correspondent value of the target function; if the calculated value is higher by at least $\theta_{occh}$ than the previous target function value saved at step 3, go to step 3; otherwise, stop the algorithm.

By this algorithm both the hybrid pairing (i.e., 2C subset) and the AO-to-AHO transformation are optimized, thus maximizing the target function and minimizing $\|\hat{\gamma}^{loc} - \hat{\gamma}\|^2 = \|\mathbf{D}^{loc} - \mathbf{D}\|^2$. It should be noted that the value of the target function is increased at both *Step 3* and *Step 4*, so that provided that if the algorithm has not been terminated yet after the convergence check performed at Step 4, the increase of the target function after these two steps is no smaller than $\theta_{occh}$. The fact that the target function is bounded from above as $\Phi_{AHO} \leq \|\mathbf{D}\|^2$ implies that the algorithm is guaranteed to terminate after $\|\mathbf{D}\|^2 / \theta_{occh}$ steps at most.

## 2.4. The LPO construction algorithm

When AO-to-AHO transformation has been found as well as the distribution of the AHOs over 1C and 2C subsets, the set of one- and two-center LPOs are created. While 1c-LPOs coincide with AHOs of 1C subset and are thus readily available, a set of AHO mixing coefficients forming the matrices $\mathbf{v}_k$ defined by (12) are to be determined for building 2c-LPOs. These matrices can be obtained from requirement that under transformation (12) the localized 1-RDM $\gamma^{loc}(\vec{r},\vec{r}')$ expressed by (1) using LPOs

$$\gamma^{loc}(\vec{r},\vec{r}') = \sum_{i \in 1c\text{-LO}} n^{loc}_i \cdot \varphi^{loc}_i(\vec{r}) \varphi^{loc}_i(\vec{r}') + \sum_{\substack{j \in 2c\text{-LO},\\ j<p(j)}} \left(\varphi^{loc}_j(\vec{r}')\right)^T \mathbf{n}_j \varphi^{loc}_j(\vec{r}') = \\ = \sum_{i \in 1C} n^{loc}_i \cdot h_i(\vec{r}) h_i(\vec{r}') + \sum_{\substack{j \in 2C,\\ j<p(j)}} \left(\mathbf{h}_j(\vec{r}')\right)^T \mathbf{v}_j^T \mathbf{n}_j \mathbf{v}_j \mathbf{h}_j(\vec{r}') \qquad (26)$$

where we introduced a 2x2 diagonal matrix

$$\mathbf{n}_j = \begin{pmatrix} n^{loc}_j & 0 \\ 0 & n^{loc}_{p(j)} \end{pmatrix}$$

containing the $j$-th pair 2c-LO occupancies (cf. (6)), becomes equivalent to Eq. (15), i.e.,

$$\gamma^{loc}(\vec{r},\vec{r}') = \sum_{i \in 1C} (\mathbf{D})_{ii} \cdot h_i(\vec{r}) h_i(\vec{r}') + \sum_{j \in 2C} (\mathbf{D})_{j,p(j)} \cdot h_j(\vec{r}) h_{p(j)}(\vec{r}'), \qquad (27)$$

where we employed (17) for the elements of $\mathbf{D}^{loc}$ matrix. Note that the condition $j < p(j)$ in (26) ensures that each pair of 2c-LOs appears in the sum only once. If the same condition is applied in the second sum of (27), the terms of this sum can be rewritten in matrix form as

$$\gamma^{loc}(\vec{r},\vec{r}') = \sum_{i \in 1C} (\mathbf{D})_{ii} \cdot h_i(\vec{r}) h_i(\vec{r}') + \sum_{\substack{j \in 2C,\\ j<p(j)}} \left(\mathbf{h}_j(\vec{r}')\right)^T \mathbf{d}_j^{pair} \mathbf{h}_j(\vec{r}') \qquad (28)$$

by introducing a 2x2 sub-matrices

$$\mathbf{d}_j^{pair} = \begin{pmatrix} (\mathbf{D})_{jj} & (\mathbf{D})_{j,p(j)} \\ (\mathbf{D})_{p(j),j} & (\mathbf{D})_{p(j),p(j)} \end{pmatrix} \qquad (29)$$

composed of the elements of $\mathbf{D}$ corresponding to the $j$-th pair of AHOs. By comparing (28) to (26) it can be readily concluded that $\mathbf{d}_j^{pair} = \mathbf{v}_j^T \mathbf{n}_j \mathbf{v}_j$ or, using orthonormality of $\mathbf{v}_j$, that

$$\mathbf{v}_j \mathbf{d}_j^{pair} \mathbf{v}_j^T = \mathbf{n}_j.$$

The latter implies that the columns of $\mathbf{v}_j^T$ must be the eigenvectors of $\mathbf{d}_j^{pair}$ matrix and that the diagonal elements of $\mathbf{n}_j$ are their corresponding eigenvalues. This condition determines the elements of $\mathbf{v}_j$ matrices (note that $\mathbf{d}_j^{pair}$ are known as soon as AHOs and their optimal pairing have been found) and hence completes the LPOs finding procedure.

## 2.5. The Chemist's Lewis-structure picture LPOs (CLPOs)

The algorithms for constructing AHOs and LPOs presented above are designed to ensure the optimality property (4) under restrictions (10) and (11) on the orbital structures. Therefore these algorithms can be considered 'nonparametric' in a sense that no additional empirical rules or adjustable parameters (apart from numerical convergence thresholds which are not essential in this context) were used to derive them. In spite of these attractive properties, the number of LPOs produced by the algorithms is equal to the number of basis functions initially used to present an input 1-RDM and therefore is typically much higher than the number of electrons in the system as soon as the quality of the basis is beyond the minimal basis[80]. This number of localized orbitals might well appear unreasonably large from the perspective of a chemist's Lewis-structure picture, in which a single doubly-occupied localized orbital is used to describe a pair of electrons in the system. Moreover, in the Lewis-structure picture a 'lone pair' (LP) of electrons, i.e., an orbital of (nearly) double occupancy localized at a single atom rather than at a pair of atoms, is an essential element, whereas the second sum in AHO target function (18) is usually maximized if as many AHOs as possible are paired (placed into 2C subset), thus leading mostly to a 2c- rather than 1c-LOs. Nevertheless, by imposing a few additional constraints on the properties of LOs obtained in the result of minimization of $\left\| \gamma - \gamma^{loc} \right\|^2$, it proves possible to bring both the number and the properties of the produced LOs in close agreement with the expectations of the chemist's Lewis-structure picture at the cost of insignificant degradation of 1-RDM approximation error (4). A set of LOs generated in this way will be referred to as Chemist's Lewis-picture property-oriented localized orbitals or CLPOs for short.

The new requirements imposed on CLPOs in addition to constraints (10) and (11), which are not abandoned, affect only 2c-LOs and can be formulated as follows:

1) In each pair of the desired 2c-LOs the occupancy of one orbital should be close to 2.0 electrons, while the occupancy of the other should be close to zero. Accordingly, the first can be considered a bonding orbital (further abbreviated as BD) and the second – antibonding orbital (or NB for short).

2) The desired 2c-LO should only correspond to a covalent but not ionic bond. That is, if the amount of ionic character of the bond is high enough, its bonding orbital must be converted into a monoatomic lone pair (LP). In our implementation, we employ bond ionicity[35] $I_j$ defined for the $j$-th pair of 2c-LOs as

$$I_j = \frac{\left| (\mathbf{v}_j)_{11}^2 - (\mathbf{v}_j)_{12}^2 \right|}{(\mathbf{v}_j)_{11}^2 + (\mathbf{v}_j)_{12}^2} = \left| (\mathbf{v}_j)_{11}^2 - (\mathbf{v}_j)_{12}^2 \right| \leq I_{thresh} \tag{30}$$

and require that it must not be greater than the user-defined threshold value $I_{thresh}$ (by default set to 0.90 in our implementation, thus ensuring that the contribution of either AHO is no less than 5%) thus formalizing the distinction between the 'truly' covalent and 'almost ionic' bonds. In (30) we utilized an orthogonality property of $\mathbf{v}_j$ matrix, which implies that $(\mathbf{v}_j)_{11}^2 + (\mathbf{v}_j)_{12}^2 = 1$ and $\left| (\mathbf{v}_j)_{22}^2 - (\mathbf{v}_j)_{21}^2 \right| = \left| (\mathbf{v}_j)_{11}^2 - (\mathbf{v}_j)_{12}^2 \right|$.

The first of the two requirements is formalized by inequalities $\max\{n^{loc}_j, n^{loc}_{p(j)}\} > 1.0$ and $\min\{n^{loc}_j, n^{loc}_{p(j)}\} < 1.0$ (i.e., one of $n^{loc}_j$, $n^{loc}_{p(j)}$ occupancies is greater than 1.0 while the other is smaller than 1.0), which must be fulfilled simultaneously. Note that the threshold value of 1.0 electrons must not be considered here as an empirical parameter, but rather as a natural limitation on how many electrons one can consider an 'electron pair'. Although this estimate might seem rather rough for distinguishing the pairs (as compared to, e.g., the threshold value of 1.90 electrons which was used in NBO approach[39]), this threshold performs quite well as illustrated in the Results and Discussion section by comparing the properties of the obtained CPLOs with the expectations of the Lewis model. Moreover, the proposed condition not only ensures that the pair consisting of both essentially unoccupied 2c-LOs ($n^{loc}_j \approx 0$ and $n^{loc}_{p(j)} \approx 0$) will never be created, but additionally prevents creating 2c-LOs composed of two LPs ($n^{loc}_j \approx 2.0$ and $n^{loc}_{p(j)} \approx 2.0$). The 2c-LOs of the latter type were possible and even favorable from the perspective of target function (18) maximization. However, they were not in line with the expectations of the chemist's Lewis-structure picture.

In summary, the new CLPOs are defined as the orbitals which provide the best possible (in the Frobenius norm sense) approximation to the true 1-RDM by localized representation

$$\gamma^{CLPO}(\vec{r},\vec{r}\,') = \sum_{i \in 1C} n^{CLPO}{}_i \cdot \varphi^{CLPO}{}_i(\vec{r})\varphi^{CLPO}{}_i(\vec{r}\,') + \\ + \sum_{j \in BD} n^{CLPO}{}_j \cdot \varphi^{CLPO}{}_j(\vec{r})\varphi^{CLPO}{}_j(\vec{r}\,') \quad (31)$$

in which only 1c-LOs (including LPs) and only BD 2c-LOs satisfying the above formulated requirements are used.

Although most of the considerations used in constructing the AHO/LPO algorithm still hold in CLPO case, a distinctive features of $\gamma^{CLPO}$ are that it includes just one diatomic BD orbital from a corresponding 2c-LO pair and that this pair itself is formed from atomic hybrids only if these hybrids, when paired, produce the BD orbital, satisfying the ionicity condition (30), and NB orbital with occupancies above and below 1.0 electrons respectively. The set of indices of atomic hybrids satisfying these requirements will be further denoted by 2CL ('two-center Lewis') and their pairing function will be denoted by $p_L(i)$. The unpaired atomic hybrids are naturally transferred into the 1C subset defined previously. In general, the optimal composition of atomic hybrids in CLPO case (i.e., $\Theta_A$ matrices in (9)) must not be exactly the same as the composition of AHOs. Therefore, the new hybrids, which are optimal for CLPO construction, will be referred to as the Lewis atomic hybrid orbitals (LHOs) to avoid ambiguity.

The derivation similar to (13)–(16) could now be repeated in order to formulate the target function, which, when maximized, generates LHOs. However, it is more convenient to arrive at this target function in a different way. By substituting (6) into (5), which are both valid in CLPO case, we find the approximation error

$$\|\gamma - \gamma^{loc}\|^2 = \sum_i \left(n^{loc}{}_i\right)^2 - 2\sum_i n^{loc}{}_i \cdot \left(\varphi^{loc}{}_i, \hat{\gamma}\varphi^{loc}{}_i\right) + \|\gamma\|^2 = \|\gamma\|^2 - \sum_i \left(n^{loc}{}_i\right)^2 \quad (32)$$

implying that the general target function is

$$\sum_i \left(n^{loc}{}_i\right)^2 = \max. \quad (33)$$

It can be noted that the obtained expression is equivalent to the target function (18), which was used earlier in LPO/AHO case, due to (6) and the fact that $\mathbf{v}_k$ matrix in (12) is orthogonal. The present CLPO case differs only in that the summands in (33) involve only a single squared occupancy

$$(n^{CLPO}{}_i)^2 = \left(\left(\mathbf{v}^L{}_i\right)_{11} h^L{}_i + \left(\mathbf{v}^L{}_i\right)_{12} h^L{}_{p_L(i)}, \hat{\gamma}\left(\left(\mathbf{v}^L{}_i\right)_{11} h^L{}_i + \left(\mathbf{v}^L{}_i\right)_{12} h^L{}_{p_L(i)}\right)\right)^2 \quad (34)$$

per 2c-LO (and LHO) pair. The quadratic form (34) is maximized if the matrix $\mathbf{v}^L{}_i$ containing LHO-to-CLPO transformation coefficients (defined in the similar way to (12)) is formed from the transposed eigenvectors of

$$\mathbf{d}^L{}_i = \begin{pmatrix} \left(h^L{}_i, \hat{\gamma}h^L{}_i\right) & \left(h^L{}_i, \hat{\gamma}h^L{}_{p_L(i)}\right) \\ \left(h^L{}_{p_L(i)}, \hat{\gamma}h^L{}_i\right) & \left(h^L{}_{p_L(i)}, \hat{\gamma}h^L{}_{p_L(i)}\right) \end{pmatrix}$$

matrix (cf. (29)). In this case the maximum value achieved by (34) is

$$(n^{LPLO}{}_i)^2 = \left(\Lambda_{i, p_L(i)}\right)^2,$$

where

$$\Lambda_{ij} = \frac{1}{2}\left(\left(h^L{}_i, \hat{\gamma}h^L{}_i\right) + \left(h^L{}_j, \hat{\gamma}h^L{}_j\right) + \sqrt{\left(\left(h^L{}_i, \hat{\gamma}h^L{}_i\right) - \left(h^L{}_j, \hat{\gamma}h^L{}_j\right)\right)^2 + 4\left(h^L{}_i, \hat{\gamma}h^L{}_j\right)^2}\right) \quad (35)$$

is the largest eigenvalue of the 2x2 sub-matrix of 1-RDO in LHO basis corresponding to the *i*-th and *j*-th LHOs. Such sub-matrix clearly coincides with $\mathbf{d}^L{}_i$ if $j = p_L(i)$, i.e., if the *i*-th and *j*-th LHOs are paired.

With the above definitions, the LHO target function can now be formulated as

$$\Phi_{LHO} = \sum_{\mu \in 1C} \left(h^L_\mu, \hat{\gamma} h^L_\mu\right)^2 + \sum_{\substack{i \in 2CL, \\ i < p_L(i)}} \left(\Lambda_{i, p_L(i)}\right)^2 \to \max. \tag{36}$$

The complete CLPO algorithm can now be formulated as follows:

*Step 1.* Create AO-to-AHO transformation as before and use AHOs as initial guess for LHOs.

*Step 2.* Determine optimal LHO pairing by using the target function (36) subjected to constraints imposed on the composition of 2CL subset and find the optimal LHO pairing thus splitting the entire set of LHOs into the 2CL and 1C subsets.

*Step 3.* Optimize AO-to-LHO transformation by the algorithm from Appendix B with the gradient matrices using the convergence threshold $\theta_{occh}$ and compute the target function (36) using the currently available 2CL subset.

*Step 4.* Use the updated AO-to-LHO transformation to determine the optimal LHO pairing and calculate the correspondent value of target function (36); if the new value is higher by at least $\theta_{occh}$ than the value of the target function saved previously at step 3, go to step 3, and otherwise stop.

It should be noted that this algorithm differs from the AHO algorithm only in constraints imposed on possible LHO pairings. This fact justifies using AHOs as initial guess at Step 1 and further suggests that the ultimate value of the target function (36) will be only slightly lower than that of (36) therefore leading to an insignificant degradation in the created CLPO approximation to the true 1-RDM (see Results and Discussion section for numerical comparison).

After the set of LHOs has been found, the $\mathbf{d}^L_i$ matrices are formed, and BD and NB CLPOs are built from LHOs using the mixing matrices $\mathbf{v}^L_i$ formed from transposed eigenvectors of $\mathbf{d}^L_i$. The LP CLPOs are formed by selecting 1c-LHOs that have occupancies above 1.0 electrons, while the remaining 1c-LHOs are treated as 'unoccupied' or 'Rydberg'[81,82] (RY) CLPOs. In this way, the total number of CLPOs is maintained the same as the number of initial basis functions (thus allowing a linear bijective transfromation between them). However, only BD and LP CLPOs are designed to make the dominant contribution to (31) and thus referred to as the Lewis subset of CLPOs while the contribution of NB and RY CLPOs (the 'non-Lewis' subset) is usually negligible.

## 3. Computation details

In order to assess the performance of the proposed orbital localization procedures and compare the results with the chemical-intuitive expectations, a test set of 33432 small molecules composed of 2 to 12 atoms was prepared based on the data available in The PubChemQC Project database[83–85]. The PubChemQC database contains over 3.98 million of molecules and ions initially taken from PubChem Compound database[86] and optimized at B3LYP/6-31G(d) level of theory[83]. The structures corresponding to closed-shell singlet molecules with zero total charge and composed of no more than 12 atoms were selected and the 1-RDM was computed for each of these molecules at density-fitted MP2/Def2-TZVPP level of theory using PSI4 program[87] (version 1.2a1.dev781). The 1-RDMs corresponding to the Hartree-Fock SCF wavefunctions available as a byproduct during the MP2 calculation were also saved and analyzed in order to assess the influence of electron correlation on the properties of localized orbitals. The MP2 calculations involved all the electrons (no frozen core approximation was used) and the resulting density matrices corresponded to the so-called 'relaxed density'. The relaxed 1-RDM, definded as an appropriate derivative of the total energy, is usually considered preferable over the density matrix determined directly from the approximate wavefunction through definition (2) since for non-variational methods (such as MP2) the relaxed densities usually lead to more correct one-electron properties[88–93]. It is worth noting that MP2 densities were reported[94] to be in better agreement with coupled-cluster densities than the DFT densities are. Besides, the MP2 densities are not constrained to 1-RDM idempotency property[95].

The PLOs and CLPOs were obtained from the relaxed MP2 and Hartree-Fock (SCF) 1-RDMs according to the presented algorithms employing the Natural Atomic Orbitals (NAOs)[42] as initial atomic orbitals $\chi_\mu(\vec{r})$. These calculations were performed using the open-source JANPA program[58] in which the newly developed algorithms were implemented.

The following error measures were used to estimate the discrepancy between the true 1-RDM $\gamma(\vec{r},\vec{r}')$ obtained from quantum-chemical calculations and its localized approximation $\gamma^{loc}(\vec{r},\vec{r}')$ of the form (1) built from the localized orbitals:

$$\varepsilon_{loc} = \frac{\left\|\gamma - \gamma^{loc}\right\|^2}{\left\|\gamma\right\|^2} = \frac{\left\|\mathbf{D} - \mathbf{D}^{loc}\right\|^2}{\left\|\mathbf{D}\right\|^2} = 1 - \frac{\left\|\mathbf{D}^{loc}\right\|^2}{\left\|\mathbf{D}\right\|^2} \tag{37}$$

(where we took advantage of the fact that $\mathbf{D}^{loc}$ is diagonal if its elements are written in the basis of localized orbitals, and, further, employed (32)) and

$$f_L = \frac{\mathrm{tr}(\hat{\gamma}^{loc})}{\mathrm{tr}(\hat{\gamma})} = \frac{\mathrm{tr}\,\mathbf{D}^{loc}}{\mathrm{tr}\,\mathbf{D}}. \tag{38}$$

The error measure $\varepsilon_{loc}$ ($0 \leq \varepsilon_{loc} \leq 1$) is the one for which the proposed orbital localization algorithms were optimized. It evaluates the overall accuracy of the localized approximation of the true 1-RDM. In contrast to that, the error measure $f_L$ ($0 \leq f_L \leq 1$) was not used in the algorithms construction. In case if the whole set of localized orbitals is used in localized 1-RDM expansion, which defines the elements of $\mathbf{D}^{loc}$ matrix, $f_L = 1$ due to the invariance of the matrix trace under the unitary transformations of orbitals. However, if the set of localized orbitals used to present $\gamma^{loc}(\vec{r},\vec{r}')$ in (1) is limited to, say, the Lewis subset (i.e., BD and LP CLPOs), $f_L < 1$ and equals the fraction of electron charge accumulated by the subset of localized Lewis orbitals. If the localized orbitals in (1) provided exact expansion for the true 1-RDM, $\varepsilon_{loc}$ would be zero and $f_L$ would be 1.0.

## 4. Results and discussion

The error measures $\varepsilon_{loc}$ and $f_L$ defined in (37) and (38) for each molecule of the test set have been calculated from 1-RDMs obtained at both HF and MP2 levels. Figs. 1 and 2 depict the obtained distributions of $\varepsilon_{loc}$ and $f_L$ values respectively in the range where the frequency with which these value occur is essentially non-zero. Since $f_L$ is guaranteed to be exactly 1.0 if the entire set of localized orbitals is used to build $\gamma^{loc}$ in (1) and $\mathbf{D}^{loc}$ in (38), Fig. 2 shows $f_L$ only for the cases when the Lewis subset (i.e., BD and LP orbitals) of the localized orbitals was used in $\mathbf{D}^{loc}$.

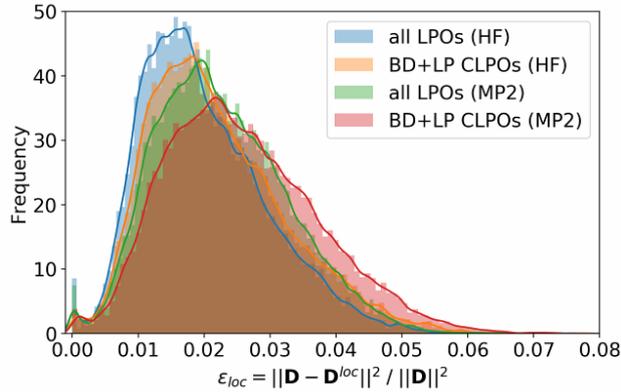
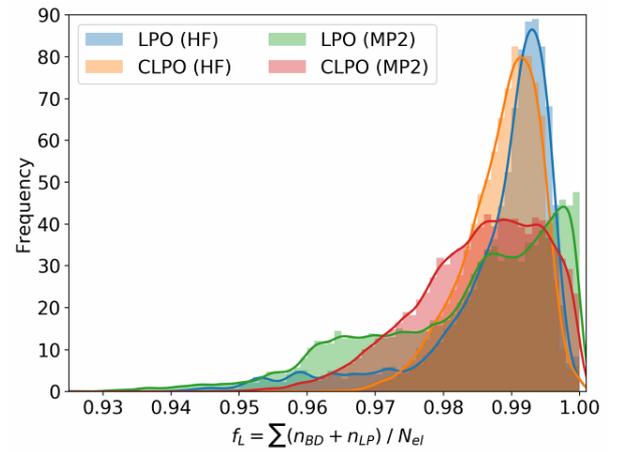

Fig. 1. Normalized distribution of relative error $\varepsilon_{loc}$ characterizing the localized approximation of 1-RDM using the complete set of LPOs ('all LPOs') and the Lewis-like subset of CLPOs ('BD+LP CLPOs')

Fig. 2. Normalized distribution of the electron charge fraction accumulated by the BD and LP localized orbitals comprising the Lewis subset of LPOs and CLPOs

The presented data show that reduction of the complete set of LPOs in (1)-type expansion of $\gamma^{loc}$ to the Lewis subset of CLPOs increases the 1-RDM approximation error insignificantly at both HF and MP2 levels. Even in the most unfavorable case, when only Lewis subset of CLPO is used to approximate 1-RDM obtained at MP2 level at which a more delocalized electronic structure is typically obtained, the overall relative error $\varepsilon_{loc}$ is mostly found in the range below 0.07 and the fraction of electronic charge accumulated by the BD and LP orbitals $f_L$ is typically well above 95%. Remarkably, the distributions of $f_L$ are concentrated in essentially the same range for LPO and CLPO. This indicates that although the number of Lewis orbitals is much higher in case of LPOs than in case of CLPOs, in fact most of LPOs have negligible occupancy (cf. Fig. 3), and thus they can be safely neglected. The remaining orbitals, being re-optimized and in this way converted into CLPOs, produce essentially as accurate localized approximation to the true 1-RDM as LPOs do.

However, an even more notable asset of CLPOs is their closer agreement with the chemist's Lewis-structure picture of molecular electronic structure. This can be demonstrated, in particular, by considering the joint distributions of the occupancies of NB and BD 2c-LOs obtained by the LPO and CLPO methods (Fig. 3).

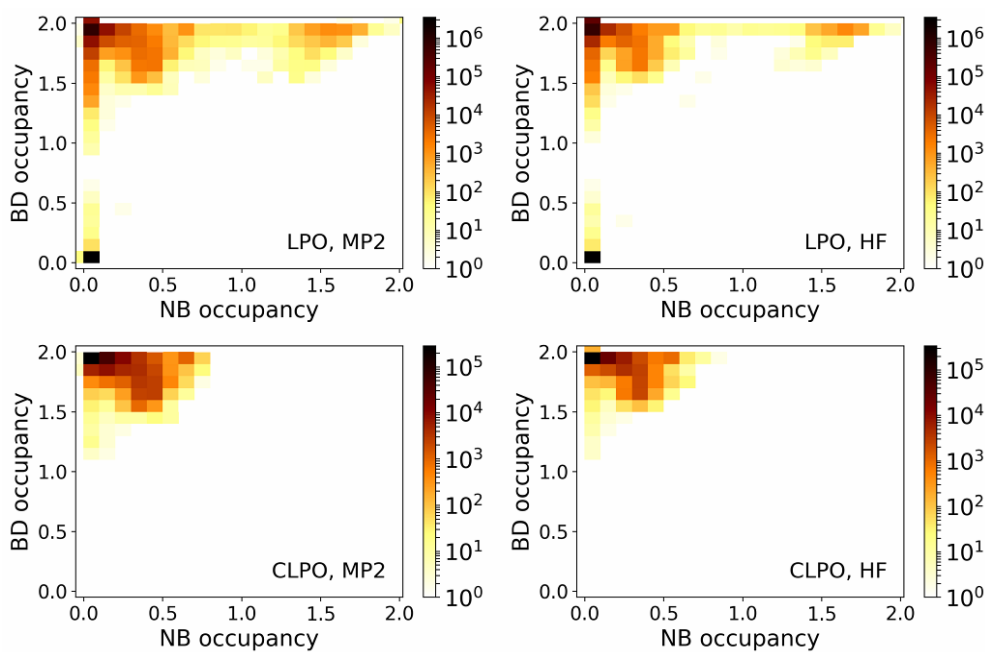

Fig. 3. Joint distributions of NB and BD orbital occupancies obtained by LPO and CLPO algorithms using MP2 and HF reduced density matrices of the test set molecules as an input. Note the logarithmic scale of color bar showing the number of samples in each cell.

In contrast to LPO NB/BD pairs exhibiting a pronounced maximum correspondent to nearly zero occupancy of both orbitals, a single maximum in CLPO case corresponds to almost doubly occupied BD orbital and almost unoccupied NB orbital (see Fig. 4 for the corresponding distribution densities). The latter is in perfect agreement with the bonding/antibonding orbital classification suggested by the Lewis-structure picture. Moreover, CLPO joint NB/BD distribution does not exhibit any maxima corresponding to simultaneously occupied NB and BD orbitals. Such maximum is, however, observed (although not so pronouncedly) for LPO pairs.

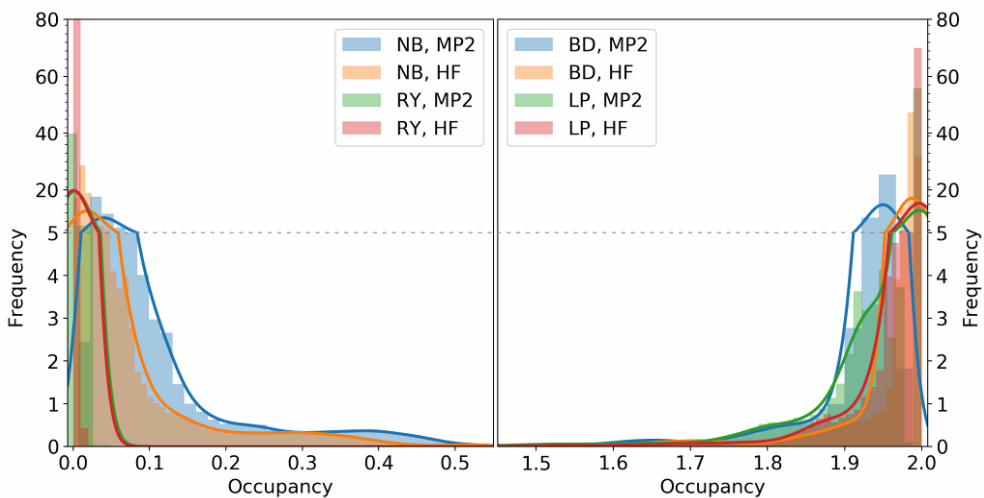

Fig. 4. Normalized distributions of occupancies of diatomic bonding (BD), antibonding (NP) as well as of monoatomic lone pair (LP) and unoccupied (RY) orbitals obtained by CLPO method using MP2 and HF reduced density matrices of the test set molecules as an input. Note a double-range linear scale on the vertical axis.

The occupancies of CLPOs are contained in rather narrow ranges corresponding to the Lewis (BD and LP with occupancies mostly above 1.7 electrons) and non-Lewis (NB and RY orbitals with occupancies mostly below 0.5 electrons) subsets, as can be seen from Fig. 4. It is remarkable that the boundaries of these ranges are much narrower than the limiting value of 1.0 electrons initially used in the CLPO algorithm for pre-selecting the atomic hybrid orbital suitable for pairing and 2c-LO formation.

The similar conclusion holds for the threshold value of $I_{thresh} = 0.90$ used to discriminate between the 'truly' covalent and almost ionic bonds in CLPO algorithm.

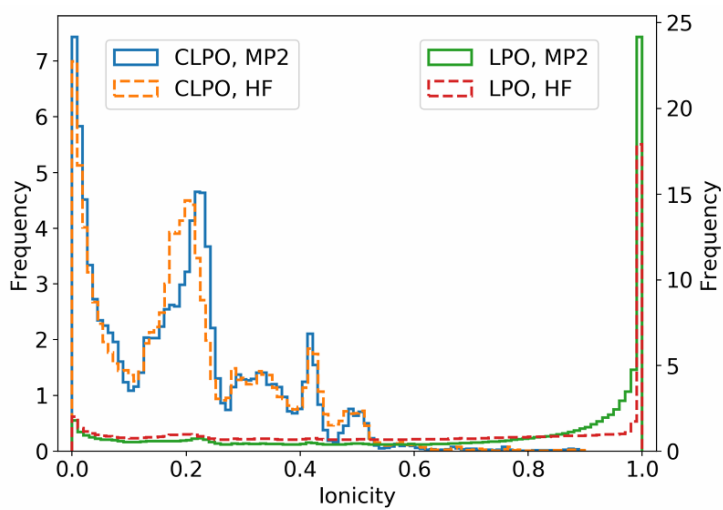

Fig. 5. Normalized distributions occupancies of bonding orbital ionicities obtained by LPO and CLPO methods using MP2 (solid lines) and HF (dashed lines) reduced density matrices of the test set molecules as an input.

As seen from Fig. 5, the CLPO BD/NB ionicity is mostly below 0.6, i.e. far below the $I_{thresh}$ value. In contrast, the ionicity of LPO BD/NB is typically close to 1.0 indicating that most of these pairs are dominated by a contribution from a single AHO only.

Further evidence of associating CLPO BD and LP orbitals with the pairs of electrons, in the Lewis-structure picture sense, comes from the close relation between the total number of BD and LP CLPOs and half the total number of electrons in the system (see Fig. 6). It should be stressed that in CLPO algorithm, no limitations are explicitly imposed on the number of BD and LP orbitals, but instead, their number results from the fulfilment of the optimal pairing requirement, which leaves some of the atomic hybrids unpaired.

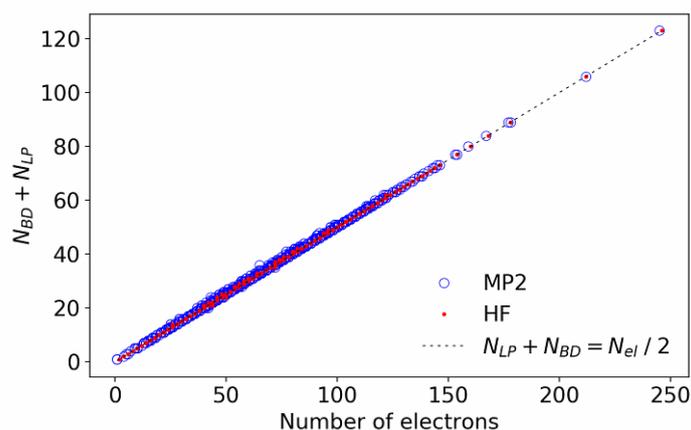

Fig. 6. The total number of Lewis electron pairs (BD and LP orbitals) obtained by the CLPO procedure using MP2 and HF reduced density matrices of the test set of molecules as an input, plotted as the function of the number of electrons in the molecule. Dashed line corresponds to half the total number of electrons in the molecule.

The BD CLPOs themselves can be safely associated with the electron pairs forming the covalent bonds. This conjecture can readily be verified by comparing the total number of BD orbitals in which atom participates with its hybridized orbitals with the typical valence of the corresponding chemical element. Such comparison has been performed for H, C, N and O atoms which are the most abundant in the molecules of the investigated test set. Fig. 7 shows the obtained distribution of the 'CLPO valencies' for these atoms.

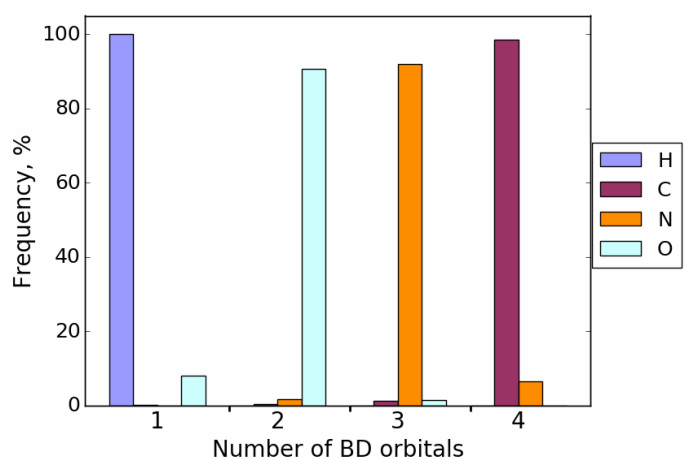

Fig. 7. Normalized distributions of the number of CLPO bonding (BD) orbitals for H, C, N, O atoms in the test set of molecules

These data demonstrate good agreement with the well-known chemical valencies of the H, C, N and O atoms. Some minor deviations observed for O and N atoms correspond to the molecules in which the electronic structure is badly representable by a single idealized Lewis structure, i.e., in which electron delocalization and resonance phenomena are pronounced.

The visual inspection of isosurfaces of BD CLPOs in the representative selection of molecules (Fig. 8) further confirms the validity of their association with covalent bond orbitals.

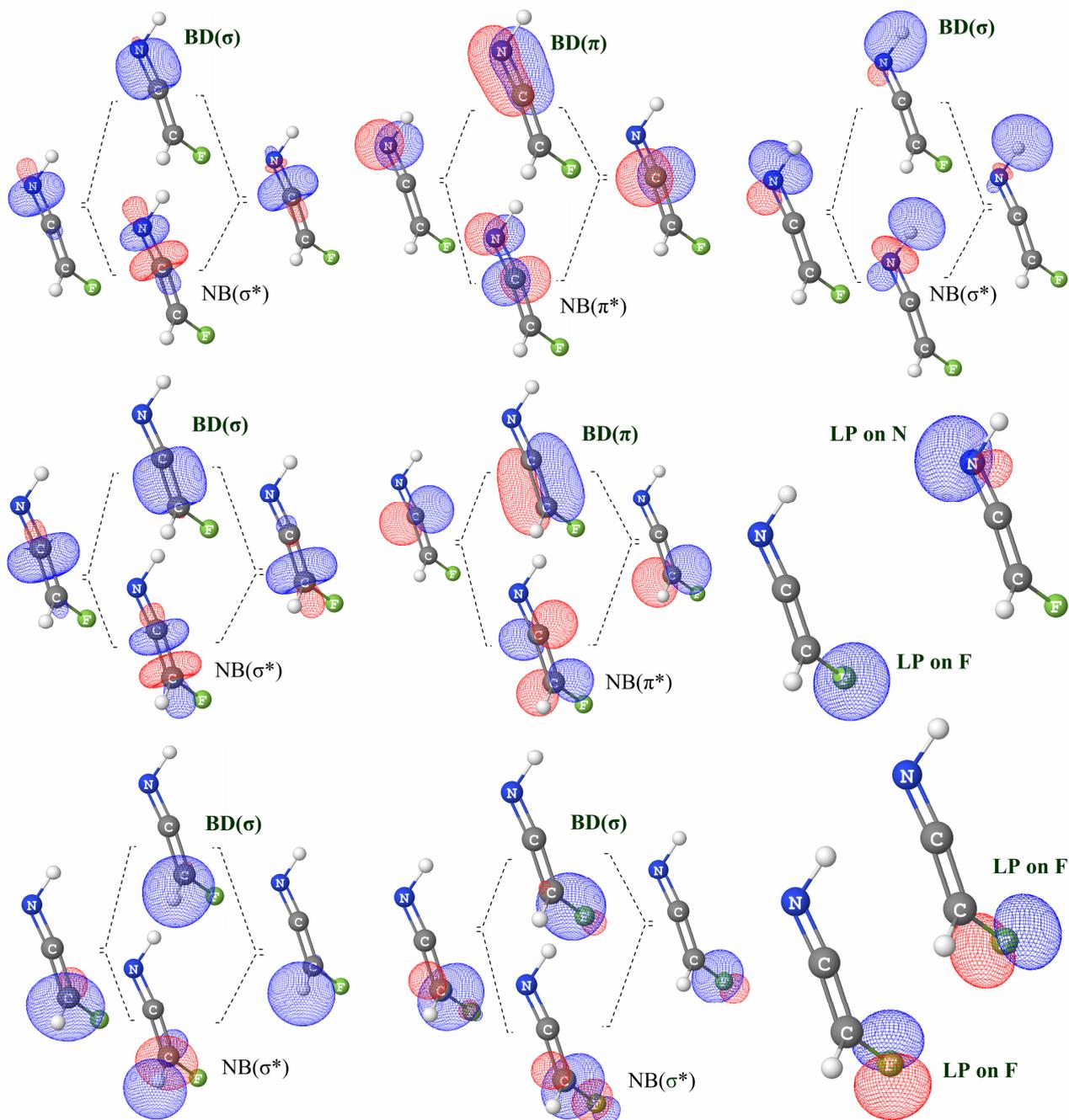

Fig. 8. CLPOs of 2-fluoroethenimine molecule and their constituent hybridized atomic orbitals (excluding the core electron orbitals) obtained by the proposed algorithms. Isosurfaces corresponding to 0.1 a.u.

## 5. Conclusions

Two procedures for obtaining the localized orbitals suitable for optimal decomposition of arbitrary one-electron molecular properties into mono- and diatomic contributions are proposed. The localized property-optimized orbitals (LPOs) are produced by the algorithm which does not involve empirical parameters and is deduced from a single optimality requirement to approximate most accurately (in a Frobenius norm sense) the first-order reduced density matrix in terms of localized orbitals of a specific structure. The chemist's Lewis-structure picture localized property-optimized orbitals (CLPOs) are derived from the similar requirement, but their bonding and lone pair orbitals subset is optimized to make the dominating contribution to the localized representation of the first-order reduced density matrix, while the antibonding and Rydberg orbitals comprise a typically minor correction. CLPO algorithm involves only a single empirical parameter which makes chemically meaningful discrimination between the covalent and highly ionic bonds. Both orbital localization procedures have been implemented in a freeware open-source program JANPA ( http://janpa.sourceforge.net/ ). The performance of the proposed procedures has been tested on a set of 33432 small closed-shell molecules containing from 2 to 12 atoms and the structure and properties of the CLPOs have been found to be in excellent agreement with the chemist's Lewis-structure picture. Thus, CLPOs can be considered as the localized orbitals forming a Lewis structure with one-electron physical properties which are closest to the properties computed from the true delocalized many-electron wavefunction.

**Appendix A**. Gradients of the AHO target function (18)

Let us introduce the density matrix $\mathbf{D}^{AO}$ by defining its elements as the coefficients of expansion of the true 1-RDM over original orthonormal AOs (cf. (13) for hybridized orbitals case):

$$\gamma(\vec{r},\vec{r}') = \sum_{\mu,\nu} \left(\mathbf{D}^{AO}\right)_{\mu\nu} \cdot \chi_\mu(\vec{r})\chi_\nu(\vec{r}') \tag{A1}$$

Due to orthonormality of AO basis we obtain $\left(\mathbf{D}^{AO}\right)_{\mu\nu} = \left(\chi_\mu, \hat{\gamma}\chi_\nu\right)$. These coefficients are usually available either directly from the quantum-chemical software used to calculate 1-RDM and/or molecular wavefunction, or can be obtained from 1-RDM in non-orthogonal basis by a linear transformation[57]. By substituting (A1) into (14) and using (9) we find

$$\left(\mathbf{D}\right)_{ij} = \sum_{\mu,\nu} \left(\mathbf{D}^{AO}\right)_{\mu\nu} \cdot \left(h_i, \chi_\mu\right)\left(\chi_\nu, h_j\right) = \sum_{\mu,\nu} \left(\mathbf{D}^{AO}\right)_{\mu\nu} \cdot \Theta_{\mu i} \Theta_{\nu j}, \tag{A2}$$

so that the density matrix in AHO basis can be expressed as

$$\mathbf{D} = \mathbf{\Theta}^T \mathbf{D}^{AO} \mathbf{\Theta},$$

where $\mathbf{\Theta}$ is the orthogonal matrix built from atomic AO-to-AHO transformation matrices $\mathbf{\Theta}_A$ (cf. (9)) as diagonal sub-blocks. Differentiation of (A2) with respect to the components of $\mathbf{\Theta}$ matrix yields

$$\frac{\partial(\mathbf{D})_{ij}}{\partial(\mathbf{\Theta})_{\alpha\beta}} = \sum_\nu \left(\mathbf{D}^{AO}\right)_{\alpha\nu} \cdot \delta_{i\beta} (\mathbf{\Theta})_{\nu j} + \sum_\mu \left(\mathbf{D}^{AO}\right)_{\mu\alpha} \cdot (\mathbf{\Theta})_{\mu i} \delta_{j\beta} =$$

$$= \left(\mathbf{D}^{AO}\mathbf{\Theta}\right)_{\alpha j} \cdot \delta_{i\beta} + \left(\mathbf{D}^{AO}\mathbf{\Theta}\right)_{\alpha i} \cdot \delta_{j\beta}, \tag{A3}$$

where we employed the fact that matrix $\mathbf{D}^{AO}$ is symmetrical.

The derivatives of the target function (18) with respect to the components of $\mathbf{\Theta}_A$ can now be obtained using the chain rule applied to the target function conveniently rewritten employing adjacency matrix $\mathbf{P}$ (see Section 2.3) as

$$\Phi_{AHO} = \sum_\mu \left((\mathbf{D})_{\mu\mu}\right)^2 + \sum_\mu \sum_{\substack{\nu, \\ \nu\neq\mu}} (\mathbf{P})_{\mu\nu} \left((\mathbf{D})_{\mu\nu}\right)^2 = \sum_{\mu,\nu} \left(\delta_{\mu\nu} + (1-\delta_{\mu\nu})(\mathbf{P})_{\mu\nu}\right)\left((\mathbf{D})_{\mu\nu}\right)^2.$$

This leads to the following result

$$\left(\mathbf{G}_{AHO}\right)_{\alpha\beta} = \frac{\partial \Phi_{AHO}}{\partial(\mathbf{\Theta})_{\alpha\beta}} = 4(\mathbf{D})_{\beta\beta}\left(\mathbf{D}^{AO}\mathbf{\Theta}\right)_{\alpha\beta} + 4\sum_\nu (1-\delta_{\beta\nu})(\mathbf{P})_{\beta\nu}(\mathbf{D})_{\beta\nu}\left(\mathbf{D}^{AO}\mathbf{\Theta}\right)_{\alpha\nu}.$$

By introducing the 'weighting matrix'

$$\left(\mathbf{W}^{AHO}\right)_{\nu\beta} = \begin{cases} (\mathbf{D})_{\beta\nu}, & \text{if } \nu \in 2C \text{ and } \nu = p(\beta) \\ (\mathbf{D})_{\beta\beta}, & \text{if } \nu = \beta \\ 0, & \text{otherwise} \end{cases}$$

the AHO target function gradient matrix can be rewritten as

$$\mathbf{G}_{AHO} = 4 \cdot \mathbf{D}^{AO} \mathbf{\Theta} \mathbf{W}^{AHO}. \tag{A4}$$

Note that since the AHO optimization algorithm involves only derivatives with respect to the elements of $\mathbf{\Theta}$ with both indices corresponding to the same atom, only diagonal sub-blocks of $\mathbf{G}_{AHO}$ matrix are to be computed.

**Appendix B**. The AO-to-AHO transformation optimization algorithm

Using the gradient matrix $\mathbf{G}_{AHO}$ (A4), the atomic hybrid optimization algorithm (cf. ref. 79) with the stepsize scaling procedure discussed in Appendix C can be formulated as follows:

*Step 1*) Set $\Phi_0 = -1$ and set $\mathbf{\Theta}^A$ to their initial approximation obtained by SDA or skip this step if some other initial values for $\mathbf{\Theta}^A$ and $\Phi_{AHO}$ are already available from the previous executions of the present optimization algorithm;

*Step 2*) Compute the target function $\Phi_{AHO}$ and sub-blocks $\mathbf{G}_A$ of its gradient matrix $\mathbf{G}_{AHO}$ for each atom using the currently available $\mathbf{\Theta}^A$ matrices;

*Step 3*) Check convergence: stop and return the accepted $\mathbf{\Theta}^A$ if: $|\Phi_{AHO} - \Phi_0| < \theta_{occh}$ and $\Phi_{AHO} > \Phi_0$, or $\Phi_{AHO} < \Phi_0$ and $\lambda \neq +\infty$ (i.e., stepsize scaling has already been applied) and $|\Phi_{AHO} - \Phi_0| < \theta_{occh}$, or if maximum number $N_{itMax}$ of iterations has been reached;

*Step 4*) If $\Phi_{AHO} > \Phi_0$, set $\Phi_0$ to $\Phi_{AHO}$ save the current $\mathbf{\Theta}^A$ as accepted, and set $\lambda$ to $+\infty$; otherwise (if $\Phi_{AHO} < \Phi_0$): discard the last values of $\mathbf{\Theta}^A$ matrices, set $\lambda$ to

$$\lambda_0 = \frac{\text{tr}(\mathbf{G}^T \mathbf{U}_0)}{\text{tr}(\mathbf{G}^T \mathbf{U}_0)} = \frac{N_{AHO}}{\sum_A \text{tr}(\mathbf{G}_A^T \mathbf{\Theta}^A)}$$ (where $N_{AHO}$ is the total number of AHOs) if $\lambda = +\infty$, or set $\lambda$ to $\lambda/2$ (i.e., reduce the current value of $\lambda$ twice) if not.

*Step 5*) Use the SVD procedure (23) to compute the new $\mathbf{\Theta}^A = \tilde{\mathbf{G}}_A (\tilde{\mathbf{G}}_A^T \tilde{\mathbf{G}}_A)^{-1/2}$, where $\tilde{\mathbf{G}}_A = \mathbf{G}_A$ if $\lambda = +\infty$, or $\tilde{\mathbf{G}}_A = \mathbf{\Theta}^A + \lambda \cdot \mathbf{G}_A$ otherwise. Proceed with Step 2.

In the current implementation both the convergence threshold $\theta_{occh}$ and the maximum number of iterations $N_{itMax}$ are user-adjustable parameters initially set to $10^{-5}$ and 1000 respectively.

**Appendix C**. Modification of Eq. (21) for adjusting the 'stepsize'

Let us assume that the linear approximation $f(\mathbf{U}) \approx f(\mathbf{U}_0) + \text{tr}(\mathbf{G}^T (\mathbf{U} - \mathbf{U}_0))$ failed, i.e., that the difference between $\mathbf{U}$ and $\mathbf{U}_0$ appeared to be too large. In this case this difference can be reduced by adding a 'penalty term' to the target function $f$ and arriving to a maximization problem for a new function

$$L(\mathbf{U}) = 2\lambda \cdot f(\mathbf{U}) - \|\mathbf{U} - \mathbf{U}_0\|^2 \tag{C1}$$

where $2\lambda$ is an adjustable parameter which controls how 'close' should $\mathbf{U}$ be to $\mathbf{U}_0$. Indeed, if the parameter $2\lambda$ is set so small that the $f(\mathbf{U})$ term can be neglected in (C1) in comparison with the 'penalty term' and $L(\mathbf{U})$ achieves its maximum value of zero at $\mathbf{U} = \mathbf{U}_0$. Further, since

$$-\|\mathbf{U}-\mathbf{U}_0\|^2 = -\mathrm{tr}\left((\mathbf{U}^T-\mathbf{U}_0^T)(\mathbf{U}-\mathbf{U}_0)\right)$$
$$= -\mathrm{tr}(\mathbf{U}^T\mathbf{U}) + \mathrm{tr}(\mathbf{U}^T\mathbf{U}_0) + \mathrm{tr}(\mathbf{U}_0^T\mathbf{U}) - \mathrm{tr}(\mathbf{U}_0^T\mathbf{U}_0),$$
$$= -2N_{dim} + 2\,\mathrm{tr}(\mathbf{U}_0^T\mathbf{U})$$

where $\mathrm{tr}(\mathbf{U}^T\mathbf{U}) = \mathrm{tr}(\mathbf{U}_0^T\mathbf{U}_0) = \mathrm{tr}\,\mathbf{1} = N_{dim}$ is the constant equal to the size of $\mathbf{U}$ and $\mathbf{U}_0$ matrices and $\left|\mathrm{tr}(\mathbf{U}_0^T\mathbf{U})\right| \leq \sqrt{\|\mathbf{U}_0\|^2} \cdot \sqrt{\|\mathbf{U}\|^2} = N_{dim}$, we conclude that the 'penalty term' is bounded from as $-4N_{dim} \leq -\|\mathbf{U}-\mathbf{U}_0\|^2 = -2N_{dim} + 2\,\mathrm{tr}(\mathbf{U}_0^T\mathbf{U}) \leq 0$. This implies that for 'large' $\lambda$, i.e., when $\lambda \cdot f(U_0) \gg 4N_{dim}$, the penalty term can be neglected and $L$ is maximized when $f(\mathbf{U})$ is. Now let us consider a case of 'intermediate' values of $\lambda$ and use a linear approximation for $L$ near $\mathbf{U}_0$:

$$L(\mathbf{U}) \approx 2\lambda \cdot f(\mathbf{U}_0) + 2\lambda \cdot \mathrm{tr}\left(\mathbf{G}^T(\mathbf{U}-\mathbf{U}_0)\right) - 2N_{dim} + 2\,\mathrm{tr}(\mathbf{U}_0^T\mathbf{U})$$
$$= L_0 + 2\,\mathrm{tr}\left((\lambda\cdot\mathbf{G}+\mathbf{U}_0)^T\mathbf{U}\right) = L_{approx}(\mathbf{U}) \quad (C2)$$

where $L_0 = 2\lambda \cdot f(\mathbf{U}_0) - 2N_{dim} - 2\lambda \cdot \mathrm{tr}(\mathbf{G}^T\mathbf{U}_0)$ does not depend on $\mathbf{U}$. The obtained function $L_{approx}(\mathbf{U})$ is maximized when

$$\mathbf{U} = \tilde{\mathbf{G}}(\tilde{\mathbf{G}}^T\tilde{\mathbf{G}})^{-1/2}, \quad (C3)$$

where

$$\tilde{\mathbf{G}} = \lambda \cdot \mathbf{G} + \mathbf{U}_0 \quad (C4)$$

is a 'damped' gradient matrix. It can be easily verified that the obtained expression (C3) reduces to $\mathbf{U} = \mathbf{U}_0$ at $\lambda = 0$ and to $\mathbf{U} = \mathbf{G}(\mathbf{G}^T\mathbf{G})^{-1/2}$ at a sufficiently large $\lambda$. Hence, (C4) can be considered as a continuous interpolation between the 'old' ('undamped') solution (21), which can violate inequality (25), and a 'fully damped' solution $\mathbf{U} = \mathbf{U}_0$ which does not violate inequality (25), but in this form is useless. It can be shown however, that apart from a special case of $\mathbf{G}^T\mathbf{U}_0$ being a symmetrical matrix, there exists a small (but finite!) value $\lambda > 0$ for which inequality (25) is fulfilled and at the same time $\mathbf{U} \neq \mathbf{U}_0$. In order to show that, consider the behavior of (C3) at small $\lambda$:

$$\mathbf{U} = (\lambda\mathbf{G}+\mathbf{U}_0)\left((\lambda\mathbf{G}+\mathbf{U}_0)^T(\lambda\mathbf{G}+\mathbf{U}_0)\right)^{-1/2} =$$
$$= (\lambda\mathbf{G}+\mathbf{U}_0)\left(\lambda^2\mathbf{G}^T\mathbf{G}+\lambda\mathbf{G}^T\mathbf{U}_0+\lambda\mathbf{U}_0^T\mathbf{G}+\mathbf{U}_0^T\mathbf{U}_0\right)^{-1/2} \approx$$
$$\approx (\lambda\mathbf{G}+\mathbf{U}_0)\left(1-\frac{\lambda}{2}(\mathbf{G}^T\mathbf{U}_0+\mathbf{U}_0^T\mathbf{G})\right) \approx$$
$$\approx \lambda\mathbf{G}+\mathbf{U}_0 - \frac{\lambda}{2}(\mathbf{U}_0\mathbf{G}^T\mathbf{U}_0+\mathbf{U}_0\mathbf{U}_0^T\mathbf{G}) = \mathbf{U}_0 + \frac{\lambda}{2}(\mathbf{G}-\mathbf{U}_0\mathbf{G}^T\mathbf{U}_0)$$

and hence,

$$\left(\frac{d}{d\lambda}\mathbf{U}(\lambda)\right)_{\lambda=0} = \frac{1}{2}(\mathbf{G}-\mathbf{U}_0\mathbf{G}^T\mathbf{U}_0).$$

It is worth noting that

$$\mathrm{tr}\left(\mathbf{G}^T\left(\frac{d\mathbf{U}}{d\lambda}\right)_{\lambda=0}\right) = \mathrm{tr}\left(\mathbf{G}^T\frac{1}{2}(\mathbf{G}-\mathbf{U}_0\mathbf{G}^T\mathbf{U}_0)\right) = \frac{1}{2}\mathrm{tr}(\mathbf{G}^T\mathbf{G}-\mathbf{G}^T\mathbf{U}_0\mathbf{G}^T\mathbf{U}_0) \quad (C5)$$

where the right-hand side can be rearranged using

$$\left\|\mathbf{GU}_0^T - \mathbf{U}_0\mathbf{G}^T\right\|^2 = \operatorname{tr}\left(\left(\mathbf{U}_0\mathbf{G}^T - \mathbf{GU}_0^T\right)\left(\mathbf{GU}_0^T - \mathbf{U}_0\mathbf{G}^T\right)\right) =$$
$$= \operatorname{tr}\left(\mathbf{U}_0\mathbf{G}^T\mathbf{GU}_0^T - \mathbf{U}_0\mathbf{G}^T\mathbf{U}_0\mathbf{G}^T - \mathbf{GU}_0^T\mathbf{GU}_0^T + \mathbf{GU}_0^T\mathbf{U}_0\mathbf{G}^T\right) = \quad (C6)$$
$$= 2\operatorname{tr}\left(\mathbf{G}^T\mathbf{G} - \mathbf{G}^T\mathbf{U}_0\mathbf{G}^T\mathbf{U}_0\right)$$

With this result we further obtain for the derivative of the target function $f$

$$\left(\frac{d}{d\lambda}f(\mathbf{U}(\lambda))\right)_{\lambda=0} = \sum_{\mu\nu}\frac{df}{dU_{\mu\nu}}\left(\frac{dU_{\mu\nu}}{d\lambda}\right)_{\lambda=0} = \operatorname{tr}\left(\mathbf{G}^T\frac{d\mathbf{U}}{d\lambda}\right) = \frac{1}{4}\left\|\mathbf{GU}_0^T - \mathbf{U}_0\mathbf{G}^T\right\|^2 \geq 0,$$

so that $\left.\frac{df}{d\lambda}\right|_{\lambda=0}$ is either zero (implying that $\mathbf{GU}_0^T = \mathbf{U}_0\mathbf{G}^T$) or positive. In the latter case there exists a *finite* range of values of the parameter $\lambda$ near zero in which the value of $f(\mathbf{U}(\lambda))$ increases as $\lambda$ increases. This, in turn, implies that one can always find such $\lambda$ which is small enough (but finite!) to fulfill inequality (25). This fact proves the validity of the step adjustment procedure using Eqs. (C3) and (C4) proposed above. Note that the case when $\left.\frac{df}{d\lambda}\right|_{\lambda=0} = 0$, i.e., $\mathbf{GU}_0^T = \mathbf{U}_0\mathbf{G}^T$, implies that $\mathbf{U}_0$ is already a locally optimal solution in a sense that no higher value of the target function can be achieved neither by an arbitrary small additive admixture of $\mathbf{G}$ to $\mathbf{U}_0$, nor by a small multiplicative modification in the form $\mathbf{U} = e^{\Xi}\mathbf{U}_0$, where $\Xi = \mathbf{X} - \mathbf{X}^H$ is a skew-Hermitian matrix which is assumed to be small ($\left\|\mathbf{U} - \mathbf{U}_0\right\|^2 = \left\|e^{\Xi} - \mathbf{1}\right\|^2 \approx \left\|\Xi\right\|^2 \ll 1$). The latter fact follows from observation that maximization of $f(e^{\Xi}\mathbf{U}_0)$ starting from initial point $\mathbf{U}_0$, which corresponds to $\Xi = \mathbf{0}$, would require increasing $\Xi$ in the direction of the corresponding gradient equal to[73] $\Gamma = \left(\frac{\partial}{\partial(\mathbf{X})_{\mu\nu}}f\left(e^{\mathbf{X}-\mathbf{X}^T}\mathbf{U}_0\right)\right)_{\mathbf{X}=\mathbf{0}} = \mathbf{GU}_0^T - \mathbf{U}_0\mathbf{G}^T$ (the Riemannian derivative[96]) which is zero if $\mathbf{GU}_0^T = \mathbf{U}_0\mathbf{G}^T$, so that no step of steepest descent can be taken in this direction.

The only clarification needed to be done for practical application of the proposed stepsize adjustment procedure is establishing a rule for selecting the value of $\lambda$ in case of violation of inequality (25). This can be accomplished by noting that since Eq. (C2) was used to derive the stepsize adjustment rules (C3)–(C4), the two $\mathbf{U}$-dependent terms it contains must be of the same order of magnitude. Indeed, if either of them dominates we immediately arrive at one of the limiting cases: a useless $\mathbf{U} = \mathbf{U}_0$ at $\lambda = 0$, or $\mathbf{U} = \mathbf{G}\left(\mathbf{G}^T\mathbf{G}\right)^{-1/2}$, possibly violating inequality (25), at large $\lambda$. We thus introduce $\lambda_0$, a characteristic order of magnitude for $\lambda$, from the requirement that $\lambda_0 \cdot \left|\operatorname{tr}\left(\mathbf{G}^T\mathbf{U}_0\right)\right| \approx \lambda_0 \cdot \left|\operatorname{tr}\left(\mathbf{G}^T\mathbf{U}\right)\right| \sim \left|\operatorname{tr}\left(\mathbf{U}_0^T\mathbf{U}\right)\right| \leq \left|\operatorname{tr}\left(\mathbf{U}_0^T\mathbf{U}_0\right)\right| = N_{dim}$, thus we set

$$\lambda_0 = \frac{N_{dim}}{\left|\operatorname{tr}\left(\mathbf{G}^T\mathbf{U}\right)\right|}.$$

The use of the upper bound for $\left|\operatorname{tr}\left(\mathbf{U}_0^T\mathbf{U}\right)\right|$ and lower bound for $\left|\operatorname{tr}\left(\mathbf{G}^T\mathbf{U}\right)\right|$ in the above estimate for $\lambda_0$ can be justified by the fact that during the search for appropriate value of $\lambda$ needed to restore the validity of inequality (25) we begin with setting $\lambda$ to $\lambda_0$ and then decrease $\lambda$ twice iteratively until inequality (25) is fulfilled. Note that due to the algorithm convergence proof outlined above the desired $\lambda$ can be chosen anywhere in a certain finite continuous range just above zero.

The generalization onto the case of target function $f(\mathbf{U}_1,...,\mathbf{U}_{N_a})$ depending on several arguments is straightforward: now an auxiliary function $L$ is introduced as

$$L(\mathbf{U}_1,...,\mathbf{U}_{N_a}) = 2\lambda \cdot f(\mathbf{U}_1,...,\mathbf{U}_{N_a}) - \sum_k \left\| \mathbf{U}_k - \mathbf{U}_k^{(0)} \right\|^2$$

and by repeating the above considerations we obtain

$$L_{approx}(\mathbf{U}_1,...,\mathbf{U}_{N_a}) = L_0 + 2\sum_k \mathrm{tr}\left( \left( \lambda \cdot \mathbf{G}_k + \mathbf{U}_k^{(0)} \right)^T \mathbf{U}_k \right)$$

implying that Eqs. (C3) and (C4) must now be used for each argument $\mathbf{U}_k$ and its appropriate partial gradient matrix $\mathbf{G}_k$. The caracteristic order of magnitude $\lambda_0$ can now be obtained from the condition

$$\lambda_0 \cdot \sum_k \left| \mathrm{tr}\left( \mathbf{G}_k^T \mathbf{U}_k^{(0)} \right) \right| \approx \lambda_0 \cdot \sum_k \left| \mathrm{tr}\left( \mathbf{G}_k^T \mathbf{U}_k \right) \right| \sim \sum_k \left| \mathrm{tr}\left( \left( \mathbf{U}_k^{(0)} \right)^T \mathbf{U}_k \right) \right| \leq \sum_k N_k ,$$

where $N_k$ is the size of the $k$-th argument matrix $\mathbf{U}_k$, yielding

$$\lambda_0 = \frac{\sum_k N_k}{\sum_k \left| \mathrm{tr}\left( \mathbf{G}_k^T \mathbf{U}_k^{(0)} \right) \right|} .$$

For the present implementation, $\sum_k N_k = N_{AHO}$, i.e., the total number of AHOs.

**References**

1. Minkin, V. I. Glossary of terms used in theoretical organic chemistry. *Pure and Applied Chemistry*, **1999,** 71.10, 1919-1981.
2. Pulay, P. Localizability of dynamic electron correlation. *Chemical physics letters*, **1983**, 100(2), 151-154.
3. Saebo, S.; Pulay, P. Local treatment of electron correlation. *Annual Review of Physical Chemistry*, **1993,** 44(1), 213-236.
4. Sparta, M.; Neese, F. Chemical applications carried out by local pair natural orbital based coupled-cluster methods. *Chem. Soc. Rev.*, **2014**, 43, 5032-5041
5. Schütz, M., Hetzer, G., & Werner, H. J. Low-order scaling local electron correlation methods. I. Linear scaling local MP2. *The Journal of chemical physics*, **1999,** 111(13), 5691-5705.
6. Liakos, D. G., Sparta, M., Kesharwani, M. K., Martin, J. M., & Neese, F. Exploring the accuracy limits of local pair natural orbital coupled-cluster theory. *Journal of chemical theory and computation*, **2015,** 11(4), 1525-1539.
7. Usvyat D, Maschio L, Schütz M. Periodic and fragment models based on the local correlation approach. *WIREs Comput Mol Sci.* **2018**; e1357. https://doi.org/10.1002/wcms.1357
8. Casassa, S., Zicovich-Wilson, C. M., & Pisani, C. Symmetry-adapted localized Wannier functions suitable for periodic local correlation methods. *Theoretical Chemistry Accounts*, **2006,** 116(4-5), 726-733.
9. O. Lombardi and M. Labarca. The ontological autonomy of the chemical world. *Foundations of Chemistry*, **2005**, 7, 125-148.
10. L. Piela. From Quantum Theory to Computational Chemistry. A Brief Account of Developments / in: *Handbook of Computational Chemistry* / Ed. Jerzy Leszczynski, Springer Netherlands, 2012, p. 1–12.
11. J. F. Ogilvie. The nature of the chemical bond—1990: There are no such things as orbitals! *Journal of Chemical Education*, **1990**, 67, 280.
12. L. Pauling. The nature of the chemical bond—1992, *J. Chem. Educ.* **1992**, 69, 519–521.

13. J.F. Ogilvie. The nature of the chemical bond 1993: There are no such things as orbitals! / in: Conceptual trends in quantum chemistry, Eds.: E. S. Kryachko, J. L. Calais, Springer Netherlands, 1994, 171-198.
14. J.J. Morwick. Should orbitals be taught in high school? *J. Chem. Educ.*, **1979**, 56, 262–263.
15. E. R. Scerri. Have orbitals really been observed?. *J. Chem. Educ.*, **2000**, 77, 1492–1494.
16. J. M. Zuo, M. O'Keefe and J. C. H. Spence. Have orbitals really been observed?. *J. Chem. Educ.*, **2001**, 78, 877.
17. E. R. Scerri. Have orbitals really been observed?. *J. Chem. Educ.*, **2002**, 79, 310.
18. Z. Jenkins. Do you need to believe in orbitals to use them?: realism and the autonomy of chemistry. *Philosophy of Science*, **2003**, 70, 1052–1062.
19. S. Shahbazian and M. Zahedi. The role of observables and non-observables in chemistry: a critique of chemical language. *Foundations of Chemistry*, **2006,** 8, 37–52.
20. K. Wittel and S. P. McGlynn. The orbital concept in molecular spectroscopy. *Chemical Reviews*, **1977**, 77, 745–771.
21. R. S. Mulliken. Spectroscopy, molecular orbitals, and chemical bonding. *Science*, **1967**, 157, 13–24.
22. G. Gryn'ova, M. L. Coote and C. Corminboeuf. Theory and practice of uncommon molecular electronic configurations., *Wiley Interdisciplinary Reviews: Computational Molecular Science*, **2015**, 5, 440–459.
23. G. Tsaparlis. Atomic orbitals, molecular orbitals and related concepts: Conceptual difficulties among chemistry students. *Research in Science Education*, **1997**, 27, 271-287.
24. G. Tsaparlis, G. Papaphotis. Quantum-chemical concepts: Are they suitable for secondary students? *Chemistry Education Research and Practice*, **2002**, 3, 129-144.
25. M. Labarca and O. Lombardi. Why orbitals do not exist? *Foundations of Chemistry*, **2010**, 12, 149-157.
26. J. Autschbach. Orbitals: some fiction and some facts. *Journal of Chemical Education,* **2012**, 89, 1032-1040.
27. F. Barradas-Solas and P.J. Sánchez Gómez. Orbitals in chemical education. An analysis through their graphical representations. *Chemistry Education Research and Practice*, **2014**, 15, 311-319.
28. D. G. Truhlar. Are molecular orbitals delocalized?. *Journal of Chemical Education,* **2012**, 89, 573–574.
29. M.Dauth, T. Körzdörfer, S. Kümmel, J. Ziroff, M. Wiessner, A. Schöll, F. Reinert, M. Arita, and K. Shimada. Orbital density reconstruction for molecules. *Physical review letters*, **2011**, 107, 193002-1–193002-5.
30. R.F.W.Bader. On the non-existence of parallel universes in chemistry. *Foundations of Chemistry,* **2011**, 13, 11–37.
31.A. Grushow Is it time to retire the hybrid atomic orbital?. *J.Chem.Educ.*, **2011**, 88, 860-862.
32. E.R. Scerri. The recently claimed observation of atomic orbitals and some related philosophical issues. *Philosophy of Science*, **2001**, 68, S76-S88.
33. P. Mulder. Are orbitals observable? *Hyle–Int J Philos Chem*, **2011,** 17, 24-35.
34. J. F. Gonthier, S. N. Steinmann, M. D. Wodrich, C. Corminboeuf*, Quantification of "fuzzy" chemical concepts: a computational perspective*, Chem. Soc. Rev. **2012**, 41, 4671-4687.
35. F. Weinhold, C. R. Landis, *Valency And Bonding: A Natural Bond Orbital Donor–Acceptor Perspective*, Cambridge University Press, New York, 2005.
36. F. Neese. Prediction of molecular properties and molecular spectroscopy with density functional theory: From fundamental theory to exchange-coupling. *Coordination Chemistry Reviews*, **2009**, 253(5-6), 526-563.
37. P.-O. Löwdin. Quantum Theory of Many-Particle Systems. I. Physical Interpretations by Means of Density Matrices, Natural Spin-Orbitals, and Convergence Problems in the Method of Configurational Interaction, *Phys. Rev.* **1955**, 97, 1474–1489.
38. F. Weinhold, *Natural bond orbital methods*, in: P.v.R. Schleyer, N.L. Allinger, T. Clark, J. Gasteiger, P.A. Kollman, H.F. Schaefer III, P.R. Schreiner (Eds.). *Encyclopedia of Computational Chemistry*, Chichester: John Wiley & Sons, **1998**; Vol. 3., pp 1792–1811.

39. A. E. Reed, L. A. Curtiss, F. Weinhold. *Intermolecular interactions from a natural bond orbital, donor-acceptor viewpoint. Chemical Reviews,* **1988**, 88(6), 899-926.
40. J. P. Foster, F. Weinhold, Natural hybrid orbitals, *J. Am. Chem. Soc.* **1980**, 102, 7211-7218.
41. A. E. Reed, F. Weinhold, Natural bond orbital analysis of near-Hartree–Fock water dimer, *J. Chem. Phys.* **1983**, 78, 4066–4073.
42. A. E. Reed, R. B. Weinstock and F. Weinhold, Natural population analysis, *J. Chem. Phys.,* **1985**, 83, 735–746.
43. F. Weinhold, J.E. Carpenter (1988) *The Natural Bond Orbital Lewis Structure Concept for Molecules, Radicals, and Radical Ions.* In: Naaman R., Vager Z. (eds) *The Structure of Small Molecules and Ions.* Springer, Boston, MA.
44. E. D. Glendening, C. R. Landis, F. Weinhold, Natural bond orbital methods, *WIREs Comput Mol Sci,* **2012**, 2, 1–42.
45. Foster, J. M., & Boys, S. Canonical configurational interaction procedure. *Reviews of Modern Physics,* **1960**, 32(2), 300–302.
46. Pipek, J., & Mezey, P. G. A fast intrinsic localization procedure applicable for abinitio and semiempirical linear combination of atomic orbital wave functions. *The Journal of Chemical Physics,* **1989**, 90(9), 4916-4926.
47. Mayer, I., Bakó, I., & Stirling, A. Are there atomic orbitals in a molecule?. *The Journal of Physical Chemistry A,* **2011**, 115(45), 12733-12737.
48. Li, Z., Li, H., Suo, B., & Liu, W. Localization of molecular orbitals: from fragments to molecule. *Accounts of chemical research,* **2014**, 47(9), 2758-2767.
49. Magnasco, V., & Perico, A. Uniform localization of atomic and molecular orbitals. I. *The Journal of Chemical Physics,* **1967**, 47(3), 971-981.
50. Jug, K. A maximum bond order principle. *Journal of the American Chemical Society,* **1977**, 99(24), 7800-7805.
51. Mayer, I. Some remarks on the maximum bond order. *Theoretical Chemistry Accounts: Theory, Computation, and Modeling (Theoretica Chimica Acta),* **1991**, 79(5), 377-378.
52. Edmiston, C., & Ruedenberg, K. Localized atomic and molecular orbitals. *Reviews of Modern Physics,* **1963**, 35(3), 457–465.
53. Edmiston, C., & Krauss, M.. Pseudonatural Orbitals as a Basis for the Superposition of Configurations. I. $He_2^+$. *J. Chem. Phys.,* **1966**, 45(5), 1833-1839.
54. Edmiston, C., & Krauss, M. Configuration-Interaction Calculation of $H_3$ and $H_2$. *J. Chem. Phys.,* **1965**, 42(3), 1119-1120.
55. Zoboki, T., & Mayer, I. Extremely localized nonorthogonal orbitals by the pairing theorem. *Journal of computational chemistry,* **2011**, 32(4), 689-695.
56. Clifford Dykstra, Gernot Frenking, Kwang Kim, Gustavo Scuseria. *Theory and applications of computational chemistry: The first forty years*-Elsevier (2005)
57. Nikolaienko, T. Yu.; Bulavin, L. A.; Hovorun, D. M. JANPA: An open source cross-platform implementation of the Natural Population Analysis on the Java platform, *Computational and Theoretical Chemistry,* **2014**, 1050, 15-22.
58. *JANPA: an open source cross-platform implementation of the Natural Population Analysis on the Java platform.* http://janpa.sf.net (Accessed Apr. 10, 2018)
59. Thakkar A. J.; Tanner A. C.; Smith V. H.. Inter-relationships between various representations of one-matrices and related densities: A road map and an example. *In* Density Matrices and Density Functionals (pp. 327-337). Springer, Dordrecht, **1987**.
60. Coleman A. J. *RDMs: How did we get here?.* In: *Many-Electron Densities and Reduced Density Matrices* (pp. 1-17), Ed. Jerzy Cioslowski. Springer, Boston, MA, **2000**.
61. Davidson E. R. Reduced Density Matrices in Quantum Chemistry (*Theoretical chemistry: a series of monographs, v. 6*); Academic Press Inc., **1976**.
62. Mayer I. *Bond orders and energy components: extracting chemical information from molecular wave functions,* CRC Press, **2017**.
63. Wiberg K. B. Application of the pople-santry-segal CNDO method to the cyclopropylcarbinyl and cyclobutyl cation and to bicyclobutane, *Tetrahedron,* **1968**, 24, 1083–1096.

64. Natiello M. A.; Medrano J. A. On the quantum theory of valence and bonding from the ab initio standpoint, *Chem. Phys. Lett.*, **1984**, 105, 180–182.
65. Kar T.; Sannigrahi A. B.; Mukherjee D. C. Comparison of atomic charges, valencies and bond orders in some hydrogen-bonded complexes calculated from Mulliken and Löwdin SCF density matrices, *J. Mol. Struct.: THEOCHEM*, **1987**, 153, 93–101.
66. Mayer I. Bond orders and valences from ab initio wave functions, *Int. J. Quantum Chem.*, **1986**, 29, 477–483.
67. Nikolaienko T. Yu.; Bulavin L. A.; Hovorun D. M. Can we treat ab initio atomic charges and bond orders as conformation-independent electronic structure descriptors? *RSC Adv.*, **2016**, 6, 74785-74796.
68. Byrne, C. L. *Signal Processing: a mathematical approach*. CRC Press, **2014**, p. 298.
69. Cardoso, J. F., & Souloumiac, A. Jacobi angles for simultaneous diagonalization. *SIAM journal on matrix analysis and applications*, **1996,** 17(1), 161-164.
70. *Multiple-View Spectral Clustering for Group-wise Functional Community Detection* https://www.mathworks.com/matlabcentral/fileexchange/58753-multiple-view-spectral-clustering-for-group-wise-functional-community-detection?focused=6695843&tab=function (Accessed: Apr., 10, 2018)
71. Edmonds, J. Maximum matching and a polyhedron with 0, 1-vertices. *Journal of Research of the National Bureau of Standards B*, **1965**, 69(125-130), 55-56.
72. Korte B., Vygen J. Maximum Matchings. In: *Combinatorial Optimization. Algorithms and Combinatorics*, vol 21. Springer, Berlin, Heidelberg, **2002**.
73. Abrudan, T. E., Eriksson, J., & Koivunen, V. Steepest descent algorithms for optimization under unitary matrix constraint. *IEEE Transactions on Signal Processing*, **2008**, 56(3), 1134-1147.
74. Keller J. B. Closest unitary, orthogonal and Hermitian operators to a given operator. *Mathematics Magazine*, **1975**, 48(4), 192-197.
75. Fan, K.; Hoffman, A. J. Some Metric Inequalities in the Space of Matrices. *Proc. Amer. Math. Soc.* **1955**, 6, 111-116.
76. Carlson B. C.; Keller J. M. Orthogonalization procedures and the localization of Wannier functions. *Physical Review* **1957**, 105(1), 102.
77. Higham, N. J. Computing the polar decomposition—with applications. *SIAM Journal on Scientific and Statistical Computing*, **1986**, 7(4), 1160-1174.
78. Philippe, B. An algorithm to improve nearly orthonormal sets of vectors on a vector processor. *SIAM Journal on Algebraic Discrete Methods*, **1987**, 8(3), 396-403.
79. Subotnik, J. E., Shao, Y., Liang, W., & Head-Gordon, M.. An efficient method for calculating maxima of homogeneous functions of orthogonal matrices: Applications to localized occupied orbitals. *The Journal of chemical physics*, **2004**, 121(19), 9220-9229.
80. Jensen, F. Atomic orbital basis sets. *Wiley Interdisciplinary Reviews: Computational Molecular Science*, **2013**, 3(3), 273-295.
81. Verhoeven, J. W. Glossary of terms used in photochemistry (IUPAC Recommendations 1996). *Pure and Applied Chemistry*, **1996**, 68(12), 2223-2286.
82. IUPAC. Compendium of Chemical Terminology, 2nd ed. (the "Gold Book"). Compiled by A. D. McNaught and A. Wilkinson. Blackwell Scientific Publications, Oxford (1997). XML on-line corrected version: http://goldbook.iupac.org (2006-) created by M. Nic, J. Jirat, B. Kosata; updates compiled by A. Jenkins. ISBN 0-9678550-9-8. https://doi.org/10.1351/goldbook.
83. Nakata M.; Shimazaki T. PubChemQC Project: a Large-Scale First-Principles Electronic Structure Database for Data-driven Chemistry, *J. Chem. Inf. Model.*, **2017**, 57 (6), 1300-1308.
84. Nakata M. The PubChemQC project: A large chemical database from the first principle calculations, *AIP Conf. Proc.* **2016**,1702, 090058.
85. *The PubChemQC Project.* http://pubchemqc.riken.jp/ (Accessed Apr. 10, 2018)
86. Kim, S.; Thiessen, P. A.; Bolton, E. E.; Chen, J.; Fu, G.; Gindulyte, A.; Han, L. Y.; He, J. E.; He, S. Q.; Shoemaker, B. A.; Wang, J. Y.; Yu, B.; Zhang, J.; Bryant, S. H.. PubChem substance and compound databases. *Nucleic acids research*, **2015**, 44(D1), D1202-D1213.
87. Parrish, R. M.; Burns, L. A.; Smith, D. G. A.; Simmonett, A. C.; DePrince III, A. E.; Hohenstein, E. G.; Bozkaya, U.; Sokolov, A. Yu.; Di Remigio, R.; Richard, R. M.; Gonthier, J. F.;

James, A. M.; McAlexander, H. R.; Kumar, A.; Saitow, M.; Wang, X.; Pritchard, B. P.; Verma, P.; Schaefer III, H. F.; Patkowski, K.; King, R. A.; Valeev, E. F.; Evangelista, F. A.; Turney, J. M.; Crawford, T. D.; Sherrill, C. D. Psi4 1.1: An Open-Source Electronic Structure Program Emphasizing Automation, Advanced Libraries, and Interoperability, *J. Chem. Theory Comput.* **2017**, 13(7), 3185–3197.

88. Diercksen, G. H.; Roos, B. O.; Sadlej, A. J. Legitimate calculation of first-order molecular properties in the case of limited CI functions. Dipole moments. *Chemical Physics*, **1981,** 59(1-2), 29-39.

89. Rendell, A. P.; Bacskay, G. B.; Hush, N. S.; Handy, N. C. The analytic configuration interaction gradient method: The calculation of one electron properties. *The Journal of chemical physics*, **1987**, 87(10), 5976-5986.

90. Packer, M. J., Dalskov, E. K., Sauer, S. P., Oddershede, J. Correlated dipole polarizabilities and dipole moments of the halides HX and $CH_3X$ (X= F, Cl and Br). *Theoretica chimica acta*, **1994,** 89(5-6), 323-333.

91. Gordon, M. S.; Schmidt, M. W.; Chaban, G. M.; Glaesemann, K. R.; Stevens, W. J.; Gonzalez, C. A natural orbital diagnostic for multiconfigurational character in correlated wave functions. *The Journal of chemical physics*, **1999,** 110(9), 4199–4207.

92. Wiberg, K.B.; Hadad, C.M.; LePage, T.J.; Breneman, C.M.; Frisch, M.J. Analysis of the effect of electron correlation on charge density distributions. *J. Phys. Chem.* **1992,** 96, 671–679

93. Tsuchimochi, T.; Ten-no, S. General technique for analytical derivatives of post-projected Hartree-Fock. *The Journal of chemical physics*, **2017,** 146(7), 074104.

94. Mezei, P. D.; Csonka, G. I.; Kállay, M. Electron density errors and density-driven exchange-correlation energy errors in approximate density functional calculations. *Journal of chemical theory and computation*, **2017,** 13(10), 4753-4764.

95. Mayer, I.; Pápai, I.; Bakó, I.; Nagy, A. Conceptual Problem with Calculating Electron Densities in Finite Basis Density Functional Theory. *Journal of chemical theory and computation*, **2017,** 13(9), 3961-3963.

96. Lehtola, S.; Jónsson, H. Unitary optimization of localized molecular orbitals. *Journal of chemical theory and computation*, **2013,** 9(12), 5365-5372.